\documentstyle[12pt,epsf]{article}

\begin{document}

\baselineskip=18pt plus 0.2pt minus 0.1pt

\makeatletter

\@addtoreset{equation}{section}
\renewcommand{\theequation}{\thesection.\arabic{equation}}
\renewcommand{\thefootnote}{\fnsymbol{footnote}}
\newcommand{\nn}{\nonumber}
\newcommand{\bm}[1]{\mbox{\boldmath $#1$}}
\newcommand{\Half}{\frac{1}{2}}
\newcommand{\half}{1/2}
\newcommand{\B}{{\cal B}} 
\newcommand{\E}{{\cal E}} 
\newcommand{\X}{X}        
\newcommand{\x}{x}        
\newcommand{\Y}{Y}        
\newcommand{\y}{y}        
\newcommand{\A}{A}        
\renewcommand{\u}{u}      
\renewcommand{\v}{v}      
\newcommand{\g}{g}        
\newcommand{\e}{e}        
\newcommand{\p}{\partial}
\newcommand{\tr}{\mathop{\rm Tr}}
\newcommand{\diag}{\mathop{\rm diag}}
\newcommand{\r}{r}
\newcommand{\hr}{\hat{\r}}
\newcommand{\bmr}{\bm{\r}}
\newcommand{\hbmr}{\hat{\bm{\r}}}
\newcommand{\Pdrv}[2]{\frac{\p #1}{\p #2}}
\newcommand{\pdrv}[2]{{\p #1}/{\p #2}}
\newcommand{\wt}[1]{\widetilde{#1}}
\newcommand{\calD}{{\cal D}}
\newcommand{\com}[2]{\left[#1,#2\right]}
\newcommand{\barm}{\overline{m}}
\newcommand{\vp}{\varphi}
\newcommand{\nr}{N}
\newcommand{\pr}{P}
\newcommand{\D}{D}
\newcommand{\Drv}[2]{\frac{d #1}{d #2}}
\newcommand{\C}[2]{{#1 \choose #2}}

\newcommand{\EBPS}{E_{\rm BPS}}

\newcommand{\vxy}[1]{\pmatrix{x_{#1}\cr y_{#1}}}
\newcommand{\veg}[1]{\pmatrix{e_{#1}\cr g_{#1}}}
\newcommand{\vuv}[1]{\pmatrix{u_{#1}\cr v_{#1}}}
\newcommand{\sgn}{\mathop{\rm sgn}}

\def\be{\begin{equation}}
\def\ee{\end{equation}}
\def\bea{\begin{eqnarray}}
\def\eea{\end{eqnarray}}
\def\ep{\epsilon}
\newcommand{\CR}{\nonumber \\}

\makeatother

\begin{titlepage}
\title{
\hfill\parbox{4cm}
{\normalsize KUNS-1503\\HE(TH)98/05 \\{\tt hep-th/9804164}}\\
\vspace{1cm}
Multi-Pronged Strings  and BPS Saturated Solutions\\
in $SU(N)$ Supersymmetric Yang-Mills Theory
}
\author{
Koji {\sc Hashimoto}\thanks{{\tt hasshan@gauge.scphys.kyoto-u.ac.jp}.
Supported in part by Grant-in-Aid for Scientific
Research from Ministry of Education, Science and Culture
(\#3160).},
{}
Hiroyuki {\sc Hata}\thanks{{\tt hata@gauge.scphys.kyoto-u.ac.jp}.
Supported in part by Grant-in-Aid for Scientific
Research from Ministry of Education, Science and Culture
(\#09640346).}
{} and
Naoki {\sc Sasakura}\thanks{{\tt sasakura@gauge.scphys.kyoto-u.ac.jp}.
Supported in part by Grant-in-Aid for Scientific
Research from Ministry of Education, Science and Culture
(\#10740119).}
\\[7pt]
{\it Department of Physics, Kyoto University, Kyoto 606-8502, Japan}
}
\date{\normalsize April, 1998}
\maketitle
\thispagestyle{empty}

\begin{abstract}
\normalsize
Extending our previous work on  $SU(3)$,
we construct spherically symmetric
BPS saturated regular configurations of ${\cal N}=4$
$SU(N)$ supersymmetric Yang-Mills theory preserving 1/4
supersymmetry, and investigate their features.
We also give exact solutions in the case some of the free parameters
of the general solutions take certain values.
These field theory BPS states correspond to the string theory
BPS states of multi-pronged strings connecting $N$ different D3-branes
by regarding the ${\cal N}=4$ supersymmetric Yang-Mills theory as an
effective field theory on parallel D3-branes.
We compare our solutions with multi-pronged strings in string picture.
\end{abstract}
\end{titlepage}


\section{Introduction}
\label{sec:intro}

In  recent developments of non-perturbative string theory,
the existence of many kinds of extended BPS objects has been realized.
These BPS states are related to each other by several duality
transformations, which are believed to be the true symmetries of
the non-perturbative string theory.
By studying a supersymmetric gauge theory
as an effective field theory on an appropriate brane configuration,
the existence of some string theory BPS states on the brane configuration
implies the existence of the corresponding BPS states in the field
theory.
Thus showing explicitly the existence of such corresponding  BPS states
by using traditional field theory techniques
would be a good test to the non-perturbative implications of
string theory and would deepen the understanding of the field theory
itself.

A typical example of such interplay is given by
the 3+1 dimensional ${\cal N}=4$ $SU(N)$ supersymmetric Yang-Mills
(SYM) theory broken spontaneously to $U(1)^{N-1}$.
This theory can be studied as an effective field theory on $N$ parallel
D3-branes \cite{WIT,TSE}.
The invariance of D3-branes under the $SL(2,Z)$ duality transformation
of type IIB string theory implies the $SL(2,Z)$ duality symmetry
of the ${\cal N}=4$ SYM theory \cite{MON,SENMON}.
The $(\pm1,0)$ strings stretching between different D3-branes
preserve 1/2 supersymmetry of the D3-brane world volume and appear as
the massive W-bosons of the broken gauge symmetries in the field
theory. The $SL(2,Z)$ duality symmetry implies the
existence of other BPS states of the field theory corresponding to the
general $(p,q)$ strings with relatively prime integers $p$ and $q$.
The states with $(p,\pm1)$ are monopoles and dyons, the field
configurations of which are explicitly known in the form of the
Prasad-Sommerfield solution \cite{PS}.
The existence of the states with $q=2$ was shown by the quantization of
the collective modes of the two monopole solutions \cite{SENMON},
while the existence of the states with $q>2$ was discussed in
\cite{POR}.

The BPS objects we discuss in this paper are multi-pronged strings in
type IIB string theory and their counterparts in ${\cal N}=4$ SYM
theory. It was conjectured originally in \cite{SHW} that a
three-pronged string would exist as a stable state
under both the conditions that the forces from the
three strings should balance and that the two-from charges of the
three strings should conserve at the junction.
The BPS feature of such a junction has been shown by
several authors in both the string picture \cite{DAS,SENNET} and the
M-theory picture \cite{KROLEE}.
The relations to the field theory BPS states have been investigated in
several contexts \cite{fieBPS,BER,FAY,BERKOL}.
In \cite{BER}, a three-pronged string connecting different parallel
D3-branes was investigated. This configuration preserves
the 1/4 of the D3-brane world volume supersymmetry, and hence
it is expected that there exist corresponding 1/4 BPS states
in the ${\cal N}=4$ $SU(N)$ SYM theory with $N>2$.
The author of \cite{BER} obtained an evidence for the existence,
showing that the mass of the three-pronged string agrees with
the mass of the corresponding field theory state under the assumption
of its BPS saturation.
But there the configuration itself in the field theory was not obtained.
Such a 1/4 BPS state was shown to have non-parallel electric and
magnetic charges \cite{BER,FH}, and this fact makes it a non-trivial
issue to solve the equations of BPS saturation.

In our previous work \cite{ours}, we constructed explicitly
the corresponding BPS states in the ${\cal N}=4$ $SU(3)$ SYM theory.
We solved explicitly the Bogomol'nyi equations in the case of
non-parallel electric and magnetic charges.
We also obtained exact solutions in the case some of the free parameters
are fixed to certain values.
The asymptotic behavior of our solutions agrees with the string
picture, that is, three straight strings meet at one point and the
forces balance at the junction.
On the other hand, we pointed out that there are some differences in our
solutions from the string picture, and the behavior of our
solutions near the junction does not seem to respect the $SL(2,Z)$
duality symmetry.
This fact may be interpreted as our technical preference that
we treat the SYM theory classically, i.e. electrically.

In this paper, we continue our investigations of string junctions
from field theoretical viewpoints, and extend our analysis to
multi-pronged strings.
The main part of this paper consists of the explicit construction of the
1/4 BPS states in ${\cal N}=4$  $SU(N)$ SYM theory $(N>2)$.
We solve the Bogomol'nyi equations under the assumption of spherical
symmetry in the D3-brane world volume.
The BPS solutions we obtain correspond to the string theory
BPS states of an $N$-pronged string connecting $N$ different
D3-branes. Such multi-pronged strings
have been recently studied in the string picture in \cite{BERKOL}.

This paper is organized as follows. In section \ref{sec:SU(N)}, we
explain our strategy to solve the Bogomol'nyi equations.
We investigate some properties of our solutions,
and give exact solutions in the case some of the free parameters of
the general solutions take certain values.
In section \ref{sec:su(4)}, we
show in detail the features of our exact solutions for $SU(4)$ and
$SU(5)$ as examples. Various interesting properties of the solutions
are also discussed there.
In section \ref{sec:force},
we calculate the long-range force between two dyons in the
SYM theory, and find the consistency with the idea that
a three-pronged string can be produced as a bound state of two
strings.
Section \ref{sec:summary} contains summary and discussions.
In appendix \ref{app:BornInfeld}, we show that our solutions satisfy
the equations of motion of the non-Abelian Born-Infeld action, and
the other appendices contain several formulas used in the text.

\newpage
\section{BPS saturated SYM configurations}
\label{sec:SU(N)}

\subsection{BPS bound and Bogomol'nyi equations}
\label{sec:BPSbound}

As mentioned in sec.\ \ref{sec:intro}, we would like to generalize the
arguments of ref.\ \cite{ours} for the $3+1$ dimensional ${\cal N}=4$
SYM with gauge group $SU(3)$ to the $SU(N)$ case and construct the BPS
saturated solutions which correspond to the string theory BPS states
of multi-pronged strings connecting $N$ D3-branes.
Omitting the fermionic terms, the energy of our $SU(N)$ SYM
system reads,
\begin{eqnarray}
U=\int d^3x \Half\tr\left\{
\left(\B_i\right)^2 + \left(\E_i\right)^2
+\left(D_i \X\right)^2 +\left(D_i \Y\right)^2
+\left(D_0 \X\right)^2 +\left(D_0 \Y\right)^2
-\com{X}{Y}^2\right\} ,
\label{eq:H}
\end{eqnarray}
where $\E_i=F_{0i}$ and $\B_i=\half\epsilon_{ijk}F_{jk}$
are the electric and magnetic fields, and the covariant derivative is
defined by $D_\mu\X=\p_\mu\X -i\left[\A_\mu,\X\right]$.
We have put the Yang-Mills coupling constant equal to one, and shall
consider the case of vanishing vacuum theta angle.
In (\ref{eq:H}), we have kept only two adjoint scalars, $X$ and $Y$,
among the original six which should describe the transverse
coordinates of the D3-branes.
This is because we are interested in the multi-pronged BPS string
states preserving $1/4$ of the supersymmetry, and such string networks
must necessarily lie on a two-dimensional plane.

In this paper we assume the D-flatness $\com{X}{Y}=0$.\footnote{
Due to the D-flatness, we can diagonalize $X$ and $Y$ simultaneously
and hence consider definite D3-brane surfaces described by the
eigenvalues of $X$ and $Y$ as we shall do in later subsections.
The Bogomol'nyi equations without imposing the D-flatness are
given in refs.\ \cite{kawano,LY}
}
To derive the BPS saturation condition, we rewire the energy
(\ref{eq:H}) as \cite{FH}
\begin{eqnarray}
&&U=\int d^3x \Half\tr\left\{
\left(\E_i\cos\theta - \B_i\sin\theta - D_i\X\right)^2
+\left(\B_i\cos\theta + \E_i\sin\theta - D_i\Y\right)^2
\right.
\nn\\
&&\qquad\qquad\left.
+\left(D_0 \X\right)^2 +\left(D_0 \Y\right)^2
\right\}
+\left(Q_\X + M_\Y\right)\cos\theta
+\left(Q_\Y - M_\X\right)\sin\theta ,
\label{eq:H2}
\end{eqnarray}
where $\theta$ is an arbitrary angle, and $Q_{\X,\Y}$ and $M_{\X,\Y}$
are defined by
\begin{equation}
Q_\X= \int\! d^3x \tr\left(\E_i D_i \X\right)
=\!\! \int\limits_{\r\to\infty}\!\! dS_i \tr\left(\E_i \X\right) ,
\quad
M_\X= \int\! d^3x \tr\left(\B_i D_i \X\right)
=\!\! \int\limits_{\r\to\infty}\!\! dS_i \tr\left(\B_i \X\right) .
\label{eq:Q&M}
\end{equation}
Since each term in the integrand of (\ref{eq:H2}) is
non-negative and $\theta$ is arbitrary, we obtain the BPS bound:
\begin{equation}
U \ge \EBPS \equiv
\sqrt{\left(Q_\X + M_\Y\right)^2 + \left(Q_\Y - M_\X\right)^2} .
\label{eq:BPSbound}
\end{equation}
The lower bound in (\ref{eq:BPSbound}) is saturated when the following
conditions hold:
\begin{eqnarray}
&&D_i\X = \E_i\cos\theta - \B_i\sin\theta , \label{eq:DX=B+E}\\
&&D_i\Y = \B_i\cos\theta + \E_i\sin\theta , \label{eq:DY=B+E}\\
&&D_0\X = D_0\Y = 0 , \label{eq:D0X=0}\\
&&\com{X}{Y}=0 , \label{eq:Dflat}
\end{eqnarray}
with the angle $\theta$ given by
\begin{equation}
\sin\theta=\frac{Q_\Y - M_\X}{\EBPS} ,
\quad
\cos\theta=\frac{Q_\X + M_\Y}{\EBPS} .
\label{eq:theta}
\end{equation}
In addition to the four equations (\ref{eq:DX=B+E})--(\ref{eq:Dflat}),
we have to impose the Gauss law,
\begin{equation}
D_i\E_i = 0 , \label{eq:DE=0}
\end{equation}
since we used it in converting the volume
integration of $Q_{\X,\Y}$ (\ref{eq:Q&M}) into the surface one.
Note that eq.\ (\ref{eq:theta}) is an automatic consequence of the two
equations (\ref{eq:DX=B+E}) and (\ref{eq:DY=B+E}) and need not be
imposed independently.
In appendix \ref{sec:susy} we show that the SYM configurations
satisfying eqs.\ (\ref{eq:DX=B+E})--(\ref{eq:Dflat}) preserve $1/4$
supersymmetry.

\subsection{SYM solutions and the IIB picture of string networks}
\label{sec:string-network}

In this subsection, we shall discuss the asymptotic ($r\to\infty$)
behavior of the solutions and its relation to the IIB picture of the
string networks.
Suppose that we have a static solution
$(\A_\mu(\bmr),\X(\bmr),\Y(\bmr))$
to the equations (\ref{eq:DX=B+E})--(\ref{eq:Dflat}) and
(\ref{eq:DE=0}), and that their asymptotic forms are,
after a suitable gauge transformation, given (locally) as follows:
\begin{eqnarray}
&&\X \sim \diag(\x_a)+\frac{1}{2r}\diag(\u_a) ,
\label{eq:asX}\\
&&\Y \sim \diag(\y_a)+\frac{1}{2r}\diag(\v_a) ,
\label{eq:asY}\\
&&\E_i \sim \frac{\hr_i}{ r^2}\;\Half\diag(\e_a) ,
\label{eq:asE}\\
&&\B_i \sim \frac{\hr_i}{ r^2}\;\Half\diag(\g_a) ,
\label{eq:asB}
\end{eqnarray}
with $\hat{\bmr}\equiv \bmr/r$.
For these asymptotic behaviors, we have
\begin{equation}
Q_X=2\pi\sum_{a=1}^N e_ax_a,
\qquad
M_X=2\pi\sum_{a=1}^N g_ax_a,
\label{eq:Q&M_comp}
\end{equation}
and similarly for $Q_Y$ and $M_Y$.

Now, this solution is interpreted as representing a configuration of
$N$ D3-branes $a=1,2,\ldots,N$ at transverse coordinates $(\x_a,\y_a)$
from which a string with two-form (NS-NS and R-R) charges
\begin{figure}[htdp]
\begin{center}
\leavevmode
\epsfxsize=70mm
\epsfbox{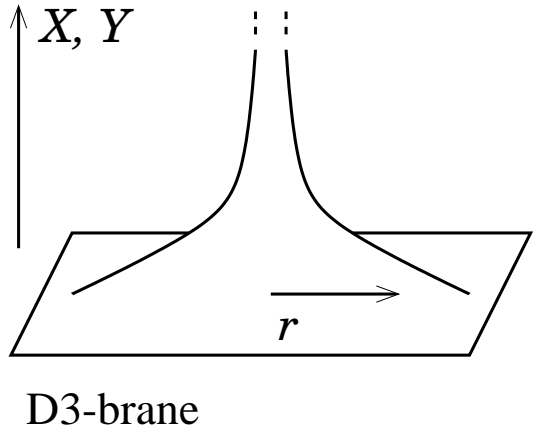}
\caption{``Tube-like'' configuration of D3-brane surface
  representing a string.
}
\label{fig:mod}
\end{center}
\end{figure}
$(\e_a,\g_a)$ are emerging in the direction $(\u_a,\v_a)$.
This is because the eigenvalues of the scalars $(\X,\Y)$ are
interpreted as the transverse coordinates of the D3-branes and
the ``tube-like'' part of the D3-brane surface (corresponding to
smaller $\r$) can be regarded as a string \cite{CH}
(see fig.\ \ref{fig:mod}).
The string directions $(\u_a,\v_a)$ are not arbitrary but are
related to $(e_a,g_a)$ by \cite{SENNET}
\begin{equation}
\pmatrix{\u_a\cr \v_a}
= -\pmatrix{\cos\theta & -\sin\theta\cr
             \sin\theta &  \cos\theta}\pmatrix{\e_a\cr \g_a} ,
\label{eq:(u,v)}
\end{equation}
where $\theta$ is the angle given by (\ref{eq:theta}).
Leaving the derivation of eq.\ (\ref{eq:(u,v)}) for our SYM
solutions in sec.\ \ref{sec:sph_symm}, we shall discuss the IIB
picture of string networks deduced from (\ref{eq:(u,v)}).

\begin{figure}[htdp]
\begin{center}
\leavevmode
\epsfxsize=120mm
\epsfbox{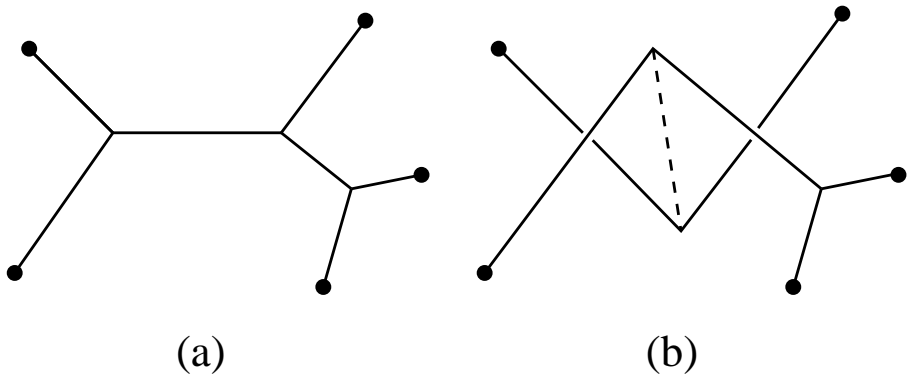}
\caption{String networks corresponding to the same set of $(e_a,g_a)$
  and $(x_a,y_a)$. The blobs represent D3-branes. The network (a) is
  physical, while (b) is unphysical since it contains a string with
  negative tension (dashed line).
}
\label{fig:network}
\end{center}
\end{figure}

In fact, we can show that, given $(e_a,g_a)$ and $(x_a,y_a)$
($a=1,2,\ldots,N$) satisfying eqs.\ (\ref{eq:theta}) and
(\ref{eq:(u,v)}), it is possible to draw ``generalized'' tree string
networks made of 3-string junctions satisfying the following two
properties:
\begin{enumerate}
\item[(A)]
The positions of the 3-string junctions in the network is consistently
given in such a way that the direction of each component
string $A$ is parallel or anti-parallel\footnote{
A string $A$ connecting two 3-string junctions (or a D3-brane
and a junction) $\alpha$ and $\beta$ with coordinates
$(x_\alpha,y_\alpha)$ and $(x_\beta,y_\beta)$ is parallel
(anti-parallel) to the vector $(u_A,v_A)$ defined to be directed from
$\alpha$ to $\beta$ if $t_A$ satisfying
$(x_\alpha,y_\alpha)+t_A(u_A,v_A)=(x_\beta,y_\beta)$ is positive
(negative).
}
to the vector $(u_A,v_A)$ which is related to the charge vector
$(e_A,g_A)$ by eq.\ (\ref{eq:(u,v)}).
Here, the component strings include the $N$ strings
emerging directly from the D3-branes as well as the internal strings
connecting 3-string junctions, and the charges $(e_A,g_A)$ are
uniquely determined by the conservation condition at the junctions.
\item[(B)]
The sum of the masses of the component strings constituting the
network coincides with the BPS bound $\EBPS$ (\ref{eq:BPSbound}) of
the energy:
\begin{equation}
\EBPS=\sum_A T_A \ell_A .
\label{eq:EBPS=sumTl}
\end{equation}
Here, $T_A=\pm 2\pi\sqrt{e_A^2 + g_A^2}$ and $\ell_A$ are
the tension and the length of the string $A$, respectively, and
the sign factor of the tension $T_A$ is positive (negative) if the
string $A$ is parallel (anti-parallel) to the vector $(u_A,v_A)$.
\end{enumerate}
In appendix \ref{app:network}, we present a proof of the properties
(A) and (B) valid for general $N$ using a technique which
reduces the problem to the simple $N=3$ case.
The string networks satisfying (A) and (B) are ``generalized'' ones
in the sense that we are allowing strings with {\em negative} tension
(see fig.\ \ref{fig:network}).
As seen from the proof given in appendix \ref{app:network}, there are
in general more than one possible string networks for a given set of
$(e_a,g_a)$ and $(x_a,g_a)$.
In order for it to be possible to identify a {\em physical} network
consisting only of strings with positive tension, $(x_a,y_a)$ cannot
be arbitrary but have to satisfy conditions determined by the charges
$(e_a,g_a)$. Such a condition is given by eq.\ (\ref{angine}) for the
$N=3$ case.
In sec.\ \ref{sec:su(4)exact}, we analyze our exact SYM solutions
satisfying the BPS saturation condition for the $N=4$ case
to find that for every parameter value of the solution there are
corresponding physical networks.
Though it is generally expected that, for $(x_a,y_a)$ and $(e_a,g_a)$
associated with a BPS saturated SYM configuration, we can always
identify physical networks, we do not yet have a proof of it.

We mentioned at the beginning of this subsection the interpretation of
the eigenvalues of the scalars $(X,Y)$ as the coordinates of the IIB
strings emerging from the D3-branes.
However, as we saw in ref.\ \cite{ours} for the $N=3$ case and shall
find later in this paper for general $N$, such interpretation does
not match the string network picture for finite $r$ since the lines
corresponding to the eigenvalues of the scalars are generally
{\em  curved}. Moreover, multiple 3-string junctions in the networks
with $N\ge 4$ is impossible in this interpretation due to the
analyticity of the scalars as functions of the four dimensional
coordinates of the SYM.


\subsection{Spherical symmetry and monopole solutions}
\label{sec:sph_symm}

Let us prepare a framework for constructing the
BPS saturated SYM configurations.
Since the angle $\theta$ can be absorbed by a rotation in the
$(\X,\Y)$ plane, we shall consider the case of $\theta=0$.
Then, the equations to be solved are
\begin{eqnarray}
&&D_i \X = \E_i , \label{eq:DX=E}\\
&&D_i \Y = \B_i , \label{eq:DY=B}
\end{eqnarray}
as well as (\ref{eq:D0X=0}), (\ref{eq:Dflat}) and (\ref{eq:DE=0}).
Our strategy for the construction of the solutions is the same as
in ref.\ \cite{ours} for the $SU(3)$ case.
First, we prepare an $SU(N)$ monopole solution
$\left(\A_i(\bmr),\Y(\bmr)\right)$ to eq.\ (\ref{eq:DY=B}).
Then, eq.\ (\ref{eq:DX=E}) is automatically satisfied by putting
$A_0(\bmr)= -\X(\bmr)$, while eq.\ (\ref{eq:D0X=0}) holds due to
eq.\ (\ref{eq:Dflat}) and the time-independence of our solution.
Therefore, we have only to solve eq.\ (\ref{eq:DE=0}), i.e.,
\begin{equation}
D_iD_i \X=0 ,
\label{eq:DDX=0}
\end{equation}
under the D-flatness condition (\ref{eq:Dflat}).
As we did in ref.\ \cite{ours}, we adopt as the monopole solutions to
eq.\ (\ref{eq:DY=B}) those given in ref.\ \cite{WB} constructed using
a general formalism for spherically symmetric solutions of ref.\
\cite{WG}. In the following, we shall present our problem in the
formalism of ref.\ \cite{WG} and explain the monopole solutions
of ref.\ \cite{WB} (see also ref.\ \cite{BW}).

Our solutions $\left(A_i, \X, \Y\right)$ are assumed to be spherically
symmetric with respect to an angular momentum operator $\bm{J}$,
i.e., they satisfy
\begin{eqnarray}
\com{J_i}{A_j}=i\epsilon_{ijk}A_k ,
\qquad
\com{J_i}{\X}=\com{J_i}{\Y}=0 ,
\label{eq:sphsymm}
\end{eqnarray}
The present $\bm{J}$ is the sum of the space and the gauge group
rotations:
\begin{equation}
\bm{J}=\bm{L} + \bm{T} ,
\label{eq:J=L+T}
\end{equation}
where $\bm{L}= -i\bmr\times\bm{\nabla}$ is the generator of the space
rotation and $\bm{T}$ is the maximal $SU(2)$ embedding in
$SU(N)$ with $T_3 =\Half\diag\left(N-1,N-3,\cdots,-N+1\right)$.
The monopole solution of \cite{WB} assumes the following form for the
vector potential:
\begin{equation}
\bm{A}(\bmr)=\left(\bm{M}(r,\hbmr) - \bm{T}\right)\times\hbmr/r ,
\label{eq:A=}
\end{equation}
where the Lie algebra valued function $M_i$ should satisfy
the spherical symmetry condition,
\begin{equation}
\com{J_i}{M_j}=i\epsilon_{ijk}M_k .
\label{eq:[J,M]}
\end{equation}
Various formulas are derived by using the expression
$\bm{\nabla}=\hbmr\pdrv{}{r}-(i/r)\hbmr\times\bm{L}$ for the space
derivative as well as the spherical symmetry properties,
eqs.\ (\ref{eq:sphsymm}) and (\ref{eq:[J,M]}).
We need in particular the following three:
\begin{eqnarray}
&&\bm{D}\Y=\hbmr \Y' + \frac{i}{r}\,\hbmr\times\com{\bm{M}}{\Y} ,
\label{eq:DY2}\\
&&\B_i = -\frac{i}{r^2}\hr_i\hr_j\left(
\Half\epsilon_{jk\ell}\com{M_k}{M_\ell} - iT_j\right)
- \frac{1}{r}\left(\hbmr\times\left(
\hbmr\times\bm{M}'\right)\right)_i ,
\label{eq:B}\\
&&D_iD_i\X = \X'' + \frac{2}{r}\X'
- \frac{1}{r^2}\left(\delta_{ij}-\hr_i\hr_j\right)
\com{M_i}{\com{M_j}{\X}} ,
\label{eq:DDX}
\end{eqnarray}
where the prime denotes the differentiation $\pdrv{}{r}$.

Due to the spherical symmetry it is sufficient to construct solutions
on the positive $z$-axis. The monopole equation (\ref{eq:DY=B}) and
the Gauss law constraint (\ref{eq:DDX=0}) on the $z$-axis are reduced
to
\begin{eqnarray}
&&r^2 \Y'= \Half\com{M_+}{M_-} - T_3 ,
\label{eq:D3Y=B3}\\
&&M'_{\pm} = \mp\com{M_\pm}{\Y} ,
\label{eq:DpmY=Bpm}\\
&&\X'' + \frac{2}{r}\X' - \frac{1}{2r^2}\Bigl(
\com{M_+}{\com{M_-}{\X}} + \com{M_-}{\com{M_+}{\X}}\Bigr)=0 ,
\label{eq:X''}
\end{eqnarray}
with $M_\pm\equiv M_1\pm i M_2$.
As seen from eqs.\ (\ref{eq:DX=E}), (\ref{eq:DY=B}) and
(\ref{eq:DY2}), the $z$-component of the electric and magnetic fields
on the positive $z$-axis are equal to the derivatives of the scalars:
\begin{equation}
\E_z = X', \qquad \B_z = Y'.
\label{eq:X'Y'}
\end{equation}

The matrix-valued functions $M_\pm(r)$ and $Y(r)$ of ref.\ \cite{WB}
are given on the positive $z$-axis as
\begin{eqnarray}
&&(M_+)_{mn} = \delta_{m,n-1}\, a_m(r) ,
\qquad M_- = M_+^T ,  \label{eq:Mpmdef}\\
&&\Y=\diag\left(Y_m(r)\right)
=\frac{1}{2r}\diag\left(\Psi_m(r) - \Psi_{m-1}(r)\right) ,
\quad \Psi_0 = \Psi_{N}=0,\label{eq:Ydef}
\end{eqnarray}
using $a_m$ and $\Psi_m$, which are further expressed in terms of
$Q_m(r)$ as
\begin{eqnarray}
&&a_m(r)=
\frac{r}{Q_m}\left(m\overline{m}Q_{m-1}Q_{m+1}\right)^{1/2} ,
\label{eq:am=Qm}\\
&&\Psi_m(r) = -r\left(\ln Q_m\right)' + m\overline{m} \qquad
(\overline{m} \equiv N - m) ,
\label{eq:Psim=Qm}
\end{eqnarray}
with $Q_0=Q_N=1$.
The functions $a_m$ and $M_\pm$ satisfy eqs.\ (\ref{eq:D3Y=B3}) and
(\ref{eq:DpmY=Bpm}) if $Q_m(r)$ is a solution to
\begin{eqnarray}
\label{eq:zenka}
  (Q_m')^2 - Q_mQ_m''=m\overline{m}\;Q_{m+1}Q_{m-1} ,
\end{eqnarray}
for $m=1,2,\ldots,N-1$.
In ref.\ \cite{WB} they found the following $Q_m$ satisfying
(\ref{eq:zenka}):
\begin{eqnarray}
Q_m = \gamma_m \sum_{D_m}\prod_{a\in D_m}e^{-2y_a r}
\prod_{b\in \overline{D}_{m}}\left(2y_b - 2y_a\right)^{-1} ,
\label{eq:sum}
\end{eqnarray}
where $\gamma_m$ is given by
\begin{eqnarray}
  \gamma_m=\frac{\prod_{n=1}^{N-1} n!}{\prod_{k=1}^{m-1}
    k!\cdot\prod_{l=1}^{\overline{m}-1} l!} ,
\end{eqnarray}
and the sum in (\ref{eq:sum}) is over the $\C{N}{m}$ distinct ways of
dividing the integers $\{1, 2,\ldots, N\}$ into two groups, $D_m$
with $m$ elements and $\overline{D}_{m}$ with $\overline{m}$
elements.

In (\ref{eq:sum}), $y_a$ ($a=1,2,\ldots,N$) are arbitrary parameters
satisfying $\sum_{a=1}^N y_a=0$.
They are nothing but $y_a$ appearing in the asymptotic expression
(\ref{eq:asY}) of the scalar $Y$ if the condition
$y_1<y_2<\cdots<y_N$ is satisfied. This is seen from the asymptotic
behavior of $\ln Q_m$,
\begin{equation}
\ln Q_m = -2\sum_{a=1}^{m}y_a\,r + \mbox{const.}
+ O\left(e^{-2(y_{m}-y_{m-1})r}\right)\quad
(r\to\infty) ,
\label{eq:asQm}
\end{equation}
and eq.\ (\ref{eq:Psim=Qm}). Using (\ref{eq:X'Y'}), we find that
the asymptotic form of the magnetic field is
$\B_z\sim -T_3/r^2$ and hence the magnetic charges $g_a$ of the
present solution are given by
\begin{equation}
\left(g_a\right) = \left(-N+1,-N+3,\ldots,N-1\right) .
\label{eq:g_a}
\end{equation}
On the other hand, the behavior of $Q_m(r)$ of (\ref{eq:sum}) near the
origin $r=0$ is
\begin{equation}
Q_m = r^{m\overline{m}}\left(1 + q_m r^2 + O(r^3)\right)
\quad (r\sim 0) ,
\label{eq:Qbeh}
\end{equation}
where the coefficient $q_m$ of the sub-leading term is
given by\footnote{
One way to derive eq.\ (\ref{eq:qm}) is as follows.
Eq.\ (\ref{eq:zenka}) implies that $q_m$ satisfies the recursion
relation $(2m\overline{m} -2)q_m = m\overline{m}(q_{m+1}-q_{m-1})$,
whose solution is given by $q_m\propto m\overline{m}$
(incidentally, this recursion equation is the $p=1$ case of eq.\
(\ref{eq:comp}), which is solved in appendix \ref{app:v}).
Taking into account the initial condition
$q_1=\frac{2}{N(N+1)}\sum_{a=1}^N y_a^2$ obtained from (\ref{eq:sum}),
we get (\ref{eq:qm}).
}
\begin{equation}
q_m = \frac{2m\overline{m}}{N(N^2-1)}\sum_{a=1}^N y_a^2 .
\label{eq:qm}
\end{equation}
The leading behavior $r^{m\overline{m}}$ of $Q_m$ is consistent with
the regularity of the solution at $r=0$.

Since we have adopted the diagonal form (\ref{eq:Ydef}) for the scalar
$Y$, the D-flatness condition (\ref{eq:Dflat}) implies that the other
scalar $X$ is also diagonal. 
The diagonal form of $X$ is also a consequence of the spherical
symmetry $\com{J_3}{X}=0$ (\ref{eq:sphsymm}) implying $\com{T_3}{X}=0$ 
on the $z$-axis.
Therefore, we express $X$ in terms of
$N-1$ functions $\Phi_m$ ($m=1,2,\ldots,N-1$) as
\begin{equation}
\X = \diag\left(\X_m\right)
\equiv\frac{1}{2r}\diag\left(\Phi_m - \Phi_{m-1}\right)
,\qquad \Phi_0 = \Phi_{N}=0.
\end{equation}
Then, eq.\ (\ref{eq:X''}) for $X$ is reduced to
\begin{equation}
\Phi_m'' - \frac{a_m^2}{r^2}\left(
2\Phi_m - \Phi_{m+1}- \Phi_{m-1} \right)=0 ,
\qquad m=1,\ldots,N-1 .
\label{eq:1}
\end{equation}
This differential equation can be solved numerically and for a certain
special values of $(y_a)$ analytically.
In the next subsection, we shall present an exact solution which is a
generalization of the exact solution for the $N=3$ case given in ref.\
\cite{ours}.

Before closing this subsection, we derive the relation
(\ref{eq:(u,v)}) for our SYM solutions. This is an immediate
consequence of (\ref{eq:X'Y'}) for $\E_z$ and $\B_z$ and the
asymptotic expressions (\ref{eq:asX}) -- (\ref{eq:asB}).
They lead to eq.\ (\ref{eq:(u,v)}) with $\theta=0$,
$(u_a,v_a)=-(e_a,g_a)$.
(Note that, since $M_\pm(r)$ decays exponentially as $r\to\infty$
as seen from eqs.\ (\ref{eq:am=Qm}) and (\ref{eq:asQm}), so do the other
components $\E_{x,y}$ and $\B_{x,y}$.)
Eq.\ (\ref{eq:(u,v)}) in the case of non-vanishing $\theta$ is
obtained by a rotation of $(X,Y)$ and hence $(u_a,v_a)$.


\subsection{Exact solutions for $X$}
\label{sec:exact_sol}

Similarly to the case of $SU(3)$ \cite{ours}, it is possible to
construct an exact solution to the equations (\ref{eq:1}) for the
following special values of $(y_a)$:
\begin{eqnarray}
\label{eq:putt}
\left(y_1,y_2,\ldots,y_N\right)
=C\left(-N\!+\!1,-N\!+\!3,\ldots,N\!-\!1\right) ,
\end{eqnarray}
where $C$ is a positive constant.
For this $(y_a)$, $Q_m$ is given explicitly by\footnote{
The easiest way to derive (\ref{eq:Qm_exact}) is to first obtain $Q_1$
from eq.\ (\ref{eq:sum}) and then use the recursion relation
(\ref{eq:zenka}) to get $Q_m$ with higher $m$.
}
\begin{eqnarray}
Q_m = \left(\frac{\sinh Cr}{C}\right)^{m\overline{m}} ,
\label{eq:Qm_exact}
\end{eqnarray}
and eq.\ (\ref{eq:1}) becomes
\begin{equation}
\Phi_m'' - \frac{m\overline{m}}{\sinh^2 r}\left(
2\Phi_m - \Phi_{m+1} - \Phi_{m-1}
\right)=0,\qquad m=1,\cdots, N\!-\!1 ,
\label{eq:eqs}
\end{equation}
where we have put $C=1$ since $C$ can be absorbed into the
rescaling of $r$.

To rewrite the differential equation (\ref{eq:eqs}) into a
diagonal form, let us consider the eigenvalue problem for the
non-differential part of (\ref{eq:eqs}):
\begin{equation}
m\overline{m}\left(2v_m^{(p)}-v_{m+1}^{(p)}-v_{m-1}^{(p)}\right)
=p(p+1) v_m^{(p)}
,\qquad m=1,\cdots, N\!-\!1 ,
\label{eq:comp}
\end{equation}
where the eigenvalue is expressed as $p(p+1)$, and we have
$v_0^{(p)}=v_N^{(p)}=0$.
The solution to eq.\ (\ref{eq:comp}) is given in appendix
\ref{app:v}. There we show that $p$ characterizing the eigenvalue
takes the integer values; $p=1,2,\ldots,N-1$. The explicit form of the
eigenvectors $v_m^{(p)}$ is also given there.
Then, by expressing $\Phi_m$ as
\begin{equation}
\Phi_m(r) = \sum_{p=1}^{N-1} v_m^{(p)}\varphi^{(p)}(r) ,
\label{eq:varphi}
\end{equation}
the differential equations (\ref{eq:eqs}) are reduced to diagonal ones
for $\varphi^{(p)}$:
\begin{equation}
\varphi^{(p)}(r)''-\frac{p(p+1)}{\sinh^2 r}\varphi^{(p)}(r)=0 .
\qquad p=1,\cdots, N\!-\!1,
\label{eq:ei}
\end{equation}

In order to solve (\ref{eq:ei}), we make a change of variables from
$r$ to $y \equiv e^{2r}$, and consider
$\wt{\varphi}^{(p)}(y)\equiv (y-1)^p\varphi^{(p)}$
instead of $\varphi^{(p)}$.
Eq.\ (\ref{eq:ei}) is then transformed into a Gauss's hypergeometric
differential equation:
\begin{eqnarray}
y(1-y)\frac{d^2}{dy^2}\wt{\varphi}^{(p)}+
\left[1-(1-2p)y \right]\frac{d}{dy}\wt{\varphi}^{(p)}
-p^2 \wt{\varphi}^{(p)}=0.
\label{eq:Gauss-f}
\end{eqnarray}
The general solution to (\ref{eq:ei}) is therefore given as a linear
combination of two independent ones:
\begin{equation}
\varphi^{(p)}=\beta_p\,(y-1)^{-p}
\Bigl(F(-p,-p,1;y) + c_p\left[
F(-p,-p,1;y)\ln y + F^*(-p,-p,1;y)\right]\Bigr)
\label{eq:solution}
\end{equation}
where\footnote{
$F^*(\alpha,\beta,\gamma;y)\equiv
\left(\pdrv{}{\alpha}+\pdrv{}{\beta}+2\pdrv{}{\gamma}
\right)F(\alpha,\beta,\gamma;y)$
}
\begin{eqnarray}
&&F(-p,-p,1;y)=\sum_{k=0}^{p}\C{p}{k}^2\;y^k ,
\label{eq:sol1}\\
&&F^*(-p,-p,1;y)= -2\sum_{k=1}^p\C{p}{k}^2
\sum_{l=1}^k\left(\frac{1}{p-l+1}+\frac{1}{l}\right)y^k ,
\label{eq:sol2}
\end{eqnarray}
and $\beta_p$ and $c_p$ are constants.
For our present purpose, $c_p$ must be chosen in such a way that
$\varphi^{(p)}$ of (\ref{eq:solution}) is non-singular at $y=1$
($r=0$). Such $c_p$ is given by
\begin{equation}
c_p=\left(2\sum_{l=1}^p\frac{1}{l}\right)^{-1} .
\label{eq:c_p}
\end{equation}
The derivation of (\ref{eq:c_p}) is given in appendix \ref{app:c_p}.
There we obtain the solution using the analytic continuation of the
hypergeometric functions.
For example, $\varphi^{(p)}$ with $p=1, 2$ and $3$ are
\begin{eqnarray}
&&\varphi^{(1)} = \beta_1\left(r\coth r-1\right),
\nn\\
&&\varphi^{(2)}
=-\beta_2\left(\coth r - r\frac{2\cosh^2 r+1}{3\sinh^2 r}\right),
\label{eq:phi(123)}\\
&&\varphi^{(3)}
=\beta_3\left(-1-\frac{15}{11\sinh^2 r}
+r\frac{3\cosh r (2\cosh^2 r+3)}{11\sinh^3 r}
\right).
\nn
\end{eqnarray}
$\varphi^{(1)}$ and $\varphi^{(2)}$ are nothing but the solutions
given in \cite{ours} as $\varphi_+$ and $\varphi_-$ for the $N=3$
case.

As for the other scalar $Y$ for the present $(y_a)$ of
(\ref{eq:putt}), $\Psi_m$ of (\ref{eq:Ydef}) is given using
(\ref{eq:Psim=Qm}) and (\ref{eq:Qm_exact}) by
\begin{equation}
\Psi_m = -m\overline{m}\left(r\coth r -1\right) .
\label{eq:Psi_exact}
\end{equation}
Since $v_m^{(1)}\propto m\overline{m}$, this is the $p=1$ term of
(\ref{eq:varphi}).


\section{Behavior of the solutions}
\label{sec:su(4)}

In this section, we shall discuss various aspects of the solutions:
explicit analysis of the exact solutions for the cases
$N=4$ and $5$ (sec.\ \ref{sec:su(4)exact}), the behavior of the
solutions near the origin for general $N$ (sec.\
\ref{sec:neartheorigin}), the concept of the effective charges (sec.\
\ref{sec:shape}), and new solutions for the cases where some of $y_a$
are degenerate (sec.\ \ref{sec:coin}).

\subsection{Exact solutions for $SU(4)$ and $SU(5)$}
\label{sec:su(4)exact}

In this subsection, we shall study the exact solutions presented in
the previous section in more detail for the cases $N=4$ and $5$.
First, let us consider the $SU(4)$ case.
Since the eigenvectors $v_m^{(1,2,3)}$ (with appropriate
normalization) read (see appendix \ref{app:v})
\begin{equation}
\vec{v}^{(1)} = \pmatrix{3\cr 4\cr 3} ,\qquad
\vec{v}^{(2)} = \pmatrix{1\cr 0\cr -1} ,\qquad
\vec{v}^{(3)} = \pmatrix{1\cr -2\cr 1} ,
\label{eq:vec}
\end{equation}
the $X$-coordinates of the four D3-branes are given by
\begin{equation}
\pmatrix{X_1\cr X_2\cr X_3\cr X_4}=
\frac{1}{2r}\pmatrix{3 & 1 &  1 \cr
                     1 & -1& -3 \cr
                     -1& -1&  3 \cr
                     -3& 1 & -1 }
\pmatrix{\varphi^{(1)}\cr \varphi^{(2)}\cr \varphi^{(3)}} ,
\label{eq:X1234}
\end{equation}
in terms of $\varphi^{(1,2,3)}$ of eq.\ (\ref{eq:phi(123)}).
The charges $(e_a,g_a)$ and the D3-brane coordinates at infinity
$(x_a,y_a)$ determined from the asymptotic form of the present exact
solution are
\begin{eqnarray}
&&(e_1,g_1) =  (3\beta_1+\beta_2+\beta_3\;,\;-3),\quad
(e_2,g_2) =  (\beta_1-\beta_2-3\beta_3\;,\;-1),\nn \\[3pt]
&&(e_3,g_3) =  (-\beta_1-\beta_2+3\beta_3\;,\;1),\quad
(e_4,g_4) =  (-3\beta_1+\beta_2-\beta_3\;,\;3),
\end{eqnarray}
and
\begin{eqnarray}
&&(x_1,y_1)=\left(\frac{3}{2}\beta_1
                     +\frac{1}{3}\beta_2
                     +\frac{3}{11}\beta_3, \frac{-3}{2}
     \right),\quad
(x_2,y_2)=\left(\frac{1}{2}\beta_1
                     -\frac{1}{3}\beta_2
                     -\frac{9}{11}\beta_3, \frac{-1}{2}
     \right),\quad
\nn \\
&&(x_3,y_3)=\left(-\frac{1}{2}\beta_1
                     -\frac{1}{3}\beta_2
                     +\frac{9}{11}\beta_3, \frac{1}{2}
     \right),\quad
(x_4,y_4)=\left(-\frac{3}{2}\beta_1
                     +\frac{1}{3}\beta_2
                     -\frac{3}{11}\beta_3, \frac{3}{2}
     \right) .
\end{eqnarray}

\begin{figure}[htdp]
\begin{center}
\leavevmode
\epsfxsize=120mm
\epsfbox{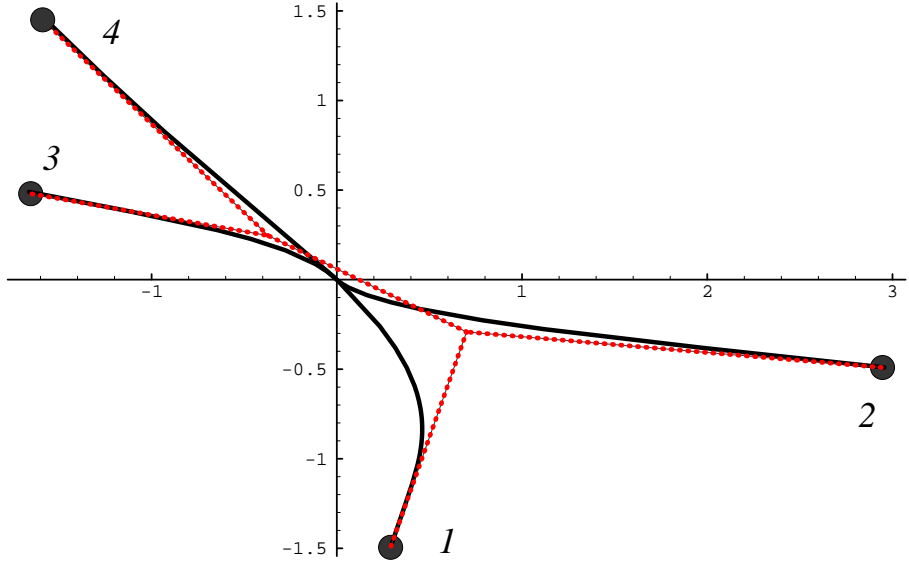}
\caption{The trajectories of $(X_a,Y_a)$ of the $N=4$ exact solution
  (solid lines). The parameters are chosen as
$(\beta_1,\beta_2,\beta_3)=(1,-2,-2)$, which give the electric and
magnetic charges $(e_a,g_a)=(-1,-3),(9,-1),(-5,1),(-3,3)$.
The dotted lines describe the corresponding four-pronged string
in the string picture.}
\label{fig:4st}
\end{center}
\end{figure}

For $(\beta_1,\beta_2,\beta_3)=(-\beta,0,0)$, we have $X_a=\beta Y_a$.
In this special case, all the D3-branes and strings are located
on a straight line which passes the origin. The charges $(e_a,g_a)$
are parallel to each other and the configuration preserves $1/2$
supersymmetries.

In fig.\ \ref{fig:4st}, an example of the typical configurations is
presented with solid curves. Notice that the four D3-brane
surfaces bend and have a common tangent at the origin. We have chosen
the parameters $\beta_p$ in order for the electric charges to take
integer values.

As explained in sec.\ \ref{sec:string-network}, we can find the
configuration of a four-pronged string in the string picture from the
\begin{figure}[htdp]
\begin{center}
\leavevmode
\epsfxsize=120mm
\epsfbox{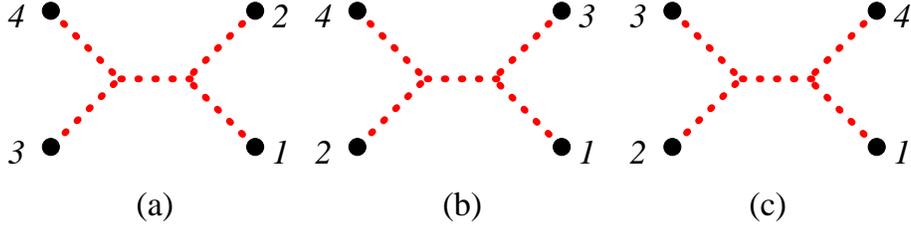}
\caption{Three different ways of connecting the four D3-branes.}
\label{fig:3case}
\end{center}
\end{figure}
asymptotic behavior of the solution. In the case of four D3-branes,
there are three different ways to connect the D3-branes by a
four-pronged string (see fig.\ \ref{fig:3case}).
Among them, the physical configurations are chosen by the condition
that the tensions of all the strings should be positive. The
four-pronged string in fig.\ \ref{fig:4st} (dotted lines)
corresponds to the case (a) of fig.\ \ref{fig:3case}.
In general, the correspondence between the parameters $\beta_p$ and
the type of the four-pronged string is as follows:
\begin{eqnarray}
\mbox{case (a)}& \Rightarrow &
|\beta_2|>5\sqrt{\frac{3}{11}}|\beta_3|
\quad\mbox{or}\quad
|\beta_2|<\frac{25}{11}|\beta_3|, \nn\\
\mbox{case (b)}&\Rightarrow &
5\sqrt{\frac{3}{11}}|\beta_3|<|\beta_2|<\frac{45}{11}|\beta_3|,
\nn\\
\mbox{case (c)}&\Rightarrow &
\frac{15}{22}|\beta_3|<|\beta_2|<5\sqrt{\frac{3}{11}}|\beta_3|.
\end{eqnarray}
Note that the whole parameter space of $\beta_p$ is covered, namely,
we can identify at least one physical four-pronged string
configurations for every value of $\beta_p$.
An interesting fact is that in the regions
\begin{equation}
5\sqrt{\frac{3}{11}}|\beta_3|<|\beta_2|<\frac{45}{11}|\beta_3| ,
\qquad
\frac{15}{22}|\beta_3|<|\beta_2|<\frac{25}{11}|\beta_3|,
\end{equation}
\begin{figure}[htdp]
\begin{center}
\leavevmode
\epsfxsize=120mm
\epsfbox{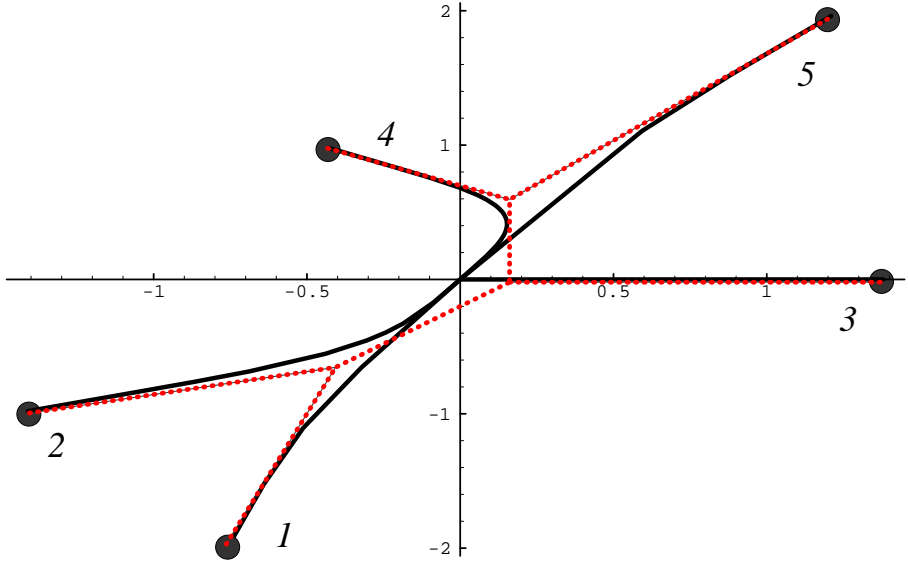}
\caption{The trajectories of the five D3-brane coordinates $(X_a,Y_a)$
  of the exact solution (solid lines), with charges $(e_a,g_a)=(-1,-4)$,
  $(-5,-2)$, $(6,0)$, $(-3,2)$, and $(3,4)$. The corresponding
  five-pronged string in the string picture is represented by dotted
  lines.}
\label{fig:5st}
\end{center}
\end{figure}
two cases satisfy the positive tension condition.
In the former region, the condition is satisfied by (a) and (b), and
in the latter by (a) and (c).
When $|\beta_2|=
5\sqrt{\frac{3}{11}}|\beta_3|$ holds,
the three cases degenerate. There is no internal string, and the four
strings meet at a point, which is not generally at the origin.

The analysis of the exact solution for $SU(5)$ can be carried out in
the same manner. In fig.\ \ref{fig:5st}, we present the trajectories
of the five D3-brane surfaces $(X_a, Y_a)$ and the corresponding
five-pronged string configuration for a certain $\beta_p$.
Notice that the D3-brane surfaces $1$, $2$, $4$ and $5$ have a common
tangent at the origin $(X,Y)=(0,0)$, while the surface 3 sticks to the
others.


\subsection{D3-brane surfaces near the origin}
\label{sec:neartheorigin}

Since the two scalars $X$ and $Y$ must vanish simultaneously at $r=0$
due to the regularity of the solution, all the $N$ D3-brane surfaces,
which are described by the eigenvalues of the scalars,
meet at the origin $(X,Y)=(0,0)$ of the transverse plane.
In this subsection we shall study how the D3-brane surfaces meet at
the origin for general $N$.

For this purpose, let us consider the behaviors near the origin of the
scalars $X$ and $Y$ in our exact solution of sec.\
\ref{sec:exact_sol}.
Rather than expanding the expression (\ref{eq:solution}) with
respect to $r$, it is easier to return to the differential equation
(\ref{eq:ei}) to know the behavior of $\varphi^{(p)}$ near $r=0$.
In fact, eq.\ (\ref{eq:ei}) is approximated near $r=0$ as
\begin{equation}
\varphi^{(p)}(r)''-\frac{p(p+1)}{r^2}\varphi^{(p)}(r)=0
\qquad (r\sim 0) ,
\label{eq:ei_r=0}
\end{equation}
and the solution regular at $r=0$ should behave as
$\varphi^{(p)}\sim r^{p+1}$. Therefore, the leading contribution to
$\Phi_m$ (\ref{eq:varphi}) comes from the $p=1$ term.
This fact, together with the expression (\ref{eq:vm1}) for $v_m^{(1)}$
implies that $\Phi_m(r)\sim 2\alpha_X m\overline{m}r^2$ ($\alpha_X$ is
an $m$-independent constant) and hence that the leading behavior of
the eigenvalue of the scalar $X$ is given by
\begin{equation}
X_m(r)\sim \alpha_X (N-2m+1) r .
\label{eq:Xmr=0}
\end{equation}
The behavior of the other scalar $Y$ should be the same as
(\ref{eq:Xmr=0}) except the constant $\alpha_X$ since $Y$ is also a
solution to the differential equation (\ref{eq:1}).
Therefore, we get
\begin{equation}
\left(X_m,Y_m\right)\sim
(N-2m+1)\left(\alpha_X, \alpha_Y\right)r
\qquad (r\sim 0) .
\label{eq:(X,Y)}
\end{equation}
The concrete expression of the constant $\alpha_Y$ is obtained
from eqs.\ (\ref{eq:Psim=Qm}) and the $q_m$ term of (\ref{eq:Qbeh}):
\begin{equation}
\alpha_Y = - \frac{2}{N(N^2-1)}\sum_a y_a^2  .
\label{eq:alpha_Y}
\end{equation}

{}From eq.\ (\ref{eq:(X,Y)}) we can deduce the followings.
The shape of the junction of $N$ D3-brane surfaces at the origin
$(X,Y)=(0,0)$ of the transverse plane differs depending on whether $N$
is even or odd.
When $N$ is even, all the $N$ D3-brane surfaces have a common
tangent at the origin: half of the D3-brane surfaces $X_m$
($1\le m\le N/2$) are ``smoothly'' connected to the other half
($N/2+1\le m\le N$) of the surfaces at the origin $(X,Y)=(0,0)$.
On the other hand, when $N$ is odd, the above picture is true except
for the D3-brane surface with $m=(N+1)/2$ since, for this particular
$m$, (\ref{eq:(X,Y)}) vanishes and the leading behavior of
$\left(X_{(N+1)/2},Y_{(N+1)/2}\right)$ is
given by the next $O(r^2)$ term which is generically not parallel to
$(\alpha_X,\alpha_Y)$.
Therefore, among $N$ D3-brane surfaces, $N-1$ have a common tangent at
the origin, while the remaining one meets with the others with an
angle.
We saw in sec.\ \ref{sec:su(4)exact} and in ref.\ \cite{ours}
that the above behaviors are actually realized
in the cases with $N=3$, $4$ and $5$ (see figs.\ \ref{fig:4st} and
\ref{fig:5st}).

The above analysis about the behavior of the scalars near $r=0$ was
concerning the exact solution of sec.\ \ref{sec:exact_sol} for $X$.
To obtain the behavior of a general solution corresponding to $(y_a)$
other than (\ref{eq:putt}), we have to analyze the differential
equation (\ref{eq:1}) near $r=0$. Naive substitution of the leading
expression of $a_m^2$, $a_m^2\sim m\overline{m}$, derived from
(\ref{eq:Qbeh}) into (\ref{eq:1}) gives eq.\ (\ref{eq:eqs}) with
$\sinh^2r$ replaced by $r^2$. Therefore, the same expansion as
(\ref{eq:varphi}) leads to (\ref{eq:ei_r=0}) and one would conclude
that the above behaviors of the scalars near the origin are not
restricted to the exact solutions.
Though we have confirmed that this naive analysis is valid for $N=3$,
$4$ and $5$, it must be checked that the other $\varphi^{(q)}$
with $q\ne p$ enters in a harmless way in the apparently higher order
terms of eq.\ (\ref{eq:ei_r=0}).

Finally in this subsection, we shall give two comments.
Our first comment is concerning the behaviors of the scalars near the
origin when $N$ is odd: one D3-brane $(X_m,Y_m)$ with $m=(N+1)/2$
stick to the rest $N-1$ branes with an angle when $N$ is odd.
Noticing that the R-R charge $g_m$ vanishes only for this particular
$m$ (c.f.\ (\ref{eq:g_a})), this phenomenon would be ascribed to the
fact that we are describing the BPS states by the classical treatment
of the SYM theory (i.e.\ electrically)
as we mentioned in ref.\ \cite{ours}.

Secondly, the fact that all the D3-brane surfaces have a common tangent
at the origin when $N$ is even means that all of them cannot be
globally straight in the transverse plane. Otherwise, the charges
$(e_a,g_a)$ are all parallel. This is the case also when $N$ is odd as
is seen from the force balance condition.

\subsection{Effective charges}
\label{sec:shape}

{}From the discussions in the previous subsection, the solutions with
non-parallel electric and magnetic charges are necessarily described
by curved lines.
As discussed in sec.\ \ref{sec:string-network} and
appendix\ \ref{app:network}, the asymptotic behavior ($r\rightarrow
\infty$) of
the curved lines reproduces the multi-pronged string configuration
in the string picture.
In this subsection, we shall point out that this feature can be extended
to any finite $r$.

The basic equations used in the proof of Appendix\ \ref{app:network}
are\footnote{We take $\theta=0$ for simplicity without losing
generality.}
\bea
\label{eq:relat1}
&&\sum_{a=1}^N\left(u_ay_a-v_ax_a\right)=0,\\
\label{eq:relat2}
&&(u_a,v_a)=-(e_a,g_a),\\
\label{eq:relat3}
&&U=|Q_X+M_Y|=2\pi\left|\sum_{a=1}^Ne_ax_a+\sum_{a=1}^Ng_ay_a\right|.
\eea
Thus if we can define quantities at finite $r$ which satisfy the
equations similar to (\ref{eq:relat1}), (\ref{eq:relat2}) and
(\ref{eq:relat3}),
then one can conclude the same results as in appendix\
\ref{app:network}, that is, we can form a multi-pronged string in the
string picture using the tangent vectors at $r$, and the configuration
has the satisfactory properties  concerning the tensions of the
strings, the BPS bound, the formation of a planer tree junction
diagram, etc.

The natural generalization of $x_a$ and $y_a$ would be just given by
the diagonal elements of the Higgs fields $X_a(r)$ and $Y_a(r)$,
where we assume a gauge in which $X(r)$ and $Y(r)$ are diagonal.
Viewing the asymptotic behaviors
(\ref{eq:asX}) and (\ref{eq:asY}), the generalization
of $(u_a,v_a)$ would be $-2r^2(X'_a(r),Y'_a(r))$,
where the prime denotes the derivative in the radial direction.
Then the generalization of (\ref{eq:relat1}) would be expected as
\begin{eqnarray}
W(r)\equiv r^2\tr\left(X'(r)Y(r)-X(r)Y'(r)\right)=0.
\label{eq:Wronskian}
\end{eqnarray}
It is enough to show (\ref{eq:Wronskian}) on the positive $z$-axis
because of the spherical symmetry.
Both $X_a$ and $Y_a$ satisfy the same differential equations
(\ref{eq:X''}), which is in the form
$(rX_a)''+\sum_{b=1}^N{A_a}^b(r)(rX_b)=0$ with
a symmetric matrix ${A_a}^b(r)$. Using this fact and taking a
derivative of (\ref{eq:Wronskian}), one finds  $W'=0$. Since $W=0$ at
$r=\infty$ from (\ref{eq:relat1}),
the equation (\ref{eq:Wronskian}) holds in the whole space.

Now let us define the $r$-dependent charges $(e_a(r),g_a(r))$ by
the electric and magnetic fields in the radial direction:
\begin{equation}
\diag\left(e_a(r)\right)= 2r^2 \E_r(r),
\qquad
\diag\left(g_a(r)\right)=2r^2\B_r(r),
\label{eq:effective_eg}
\end{equation}
where the diagonal form of the ${\cal E}_r(r)$ and ${\cal B}_r(r)$
is guaranteed by
(\ref{eq:DY2}) coming from the spherical symmetry anzats.
Then (\ref{eq:relat2}) trivially holds from the Bogomol'nyi equations
(\ref{eq:DX=E}) and (\ref{eq:DY=B}).

\begin{figure}[htdp]
\begin{center}
\leavevmode
\epsfxsize=140mm
\epsfbox{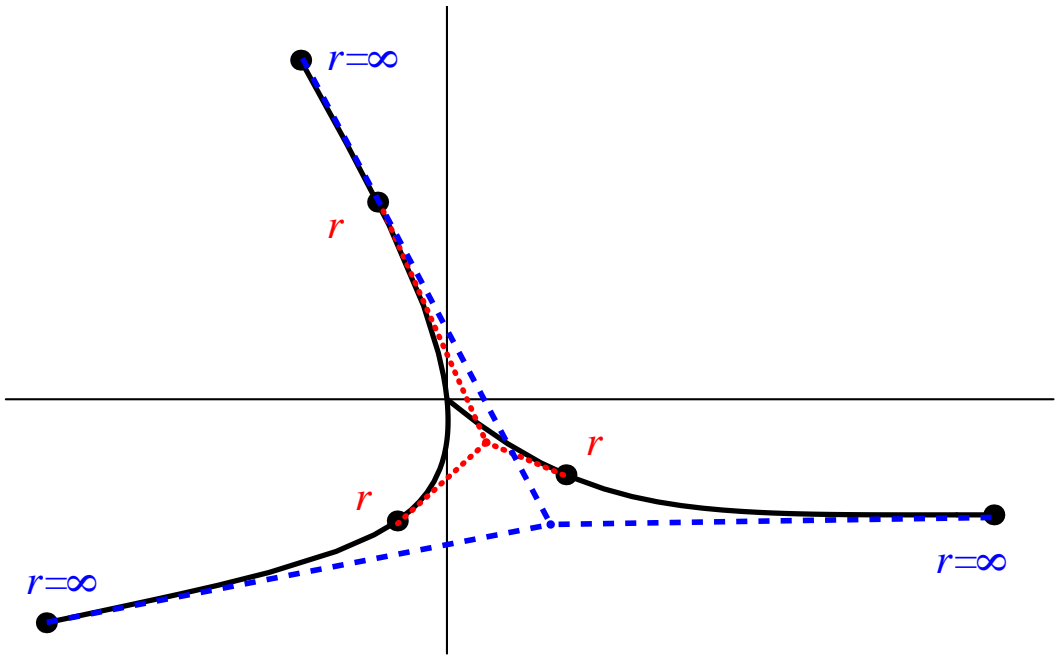}
\caption{The tangent lines to the D3-brane surfaces (solid curves) at
  $r=\infty$ (dashed lines) and at a finite $r$ (dotted lines).
}
\label{fig:shape}
\end{center}
\end{figure}

As for the last equation (\ref{eq:relat3}),
we first define the energy in the spherical region
$|x|<r$ in the D3-brane world volume by
\be
U_r=\int_{|x|<r}d^3x\  {\cal U},
\ee
where ${\cal U}$ is the energy density appearing in (\ref{eq:H2}).
Because of BPS saturation, the equality in (\ref{eq:BPSbound}) holds
for the energy $U_r$ and the charges $Q_{X,Y}(r)$ and $M_{X,Y}(r)$,
the definitions of which follow (\ref{eq:Q&M}) for the
spherical region $|x|<r$.
{}From (\ref{eq:DX=E}) and (\ref{eq:DY=B}), $Q_Y(r)-M_X(r)=0$ holds, and
hence the two charges do not contribute to the BPS bound of $U_r$.
Plugging (\ref{eq:DY2}) and the same equation
for $X$ into the definitions of the charges $Q_X(r)$ and $M_Y(r)$ and
noting the fact that the second term of (\ref{eq:DY2}) does not
contribute, they become
\be
Q_X(r)=4\pi r^2 {\rm Tr}(X(r)X'(r)),\quad
M_Y(r)=4\pi r^2 {\rm Tr}(Y(r)Y'(r)).
\ee
Thus from the definition of the effective charges
(\ref{eq:effective_eg}) and the Bogomol'nyi equations
(\ref{eq:DX=E}) and (\ref{eq:DY=B}), the same equation as
(\ref{eq:relat3}) holds for the quantities we have defined.

Because of the regularity of the solutions at $r=0$, it is a general
property that the effective charges $(e_a(r),g_a(r))$ vary as
functions of $r$ and tend to zero as $r\to 0$.
If the ratio $e_a(r):g_a(r)$ depends on $r$,
the D3-brane trajectories bend.
We show the $SU(3)$ case in fig.\ \ref{fig:shape}  as an example.
The three straight lines starting at
$(X_a(r),Y_a(r))$ in the direction $-(e_a(r),g_a(r))$ meets at a
common point. Namely, the trajectories of the D3-branes on the
transverse plane is developed in such
a way that for every $r$ the tangent lines to the trajectories
form a string junction carrying the effective charges.


\subsection{Solutions with degenerate $y_a$}
\label{sec:coin}

Up to now we have considered the cases with distinct $y_a$.
When some of the $y_a$ degenerate, we can construct solutions with
magnetic charges different from (\ref{eq:g_a}) \cite{WB}. In this
subsection, we shall discuss solutions with degenerate $y_a$ for the
$SU(4)$ case.

Since we have $y_1\!\leq y_2 \leq y_3 \leq y_4$,
the patterns of the degeneracy are classified as
\begin{equation}
 (y_1,y_2,y_3,y_4)
= \cases{
{\rm (i)}& $(-s,-s,s,s)$, \cr
{\rm (ii)} & $(-s,-s,-s,3s)$ or $(-3s,s,s,s)$, \cr
{\rm (iii)} & Only one pair in $(y_a)$ is degenerate,}
\label{eq:5cases}
\end{equation}
with $s>0$.
As stated in sec.\ \ref{sec:neartheorigin}, the analysis of the
solutions of the differential equations (\ref{eq:1}) at $r\sim0$ does
not depend on the values of $y_a$.
On the other hand, as we shall show below,
the asymptotic behaviors of the solutions
change drastically.

Let us first examine the case (i).
Since eq.\ (\ref{eq:sum}) is ill-defined for degenerate $y_a$,
we introduce an infinitesimally small parameter $\delta$ as follows:
\begin{eqnarray}
    (y_1,y_2,y_3,y_4)=
(-s+\alpha_1\delta,  -s+\alpha_2\delta,
s+\alpha_3\delta, s+\alpha_4\delta),
\end{eqnarray}
where $\sum_{a=1}^4\alpha_a=0$. Substituting
this into eq.\ (\ref{eq:sum}) and taking the limit
$\delta\rightarrow0$, we obtain the asymptotic behaviors
\begin{eqnarray}
\label{eq:asympQ}
  Q_1(r)\sim \frac{3}{8s^2}re^{2sr},\quad
  Q_2(r)\sim \frac{3}{64s^4}e^{4sr},\quad
  Q_3(r)\sim \frac{-3}{8s^2}re^{2sr},\quad
(r\to\infty) .
\end{eqnarray}
It follows from these equations that the magnetic
charges of this solution are given by $(g_a)=(-2,-2,2,2)$.
Then, substituting (\ref{eq:asympQ}) into (\ref{eq:am=Qm}),
the differential equation (\ref{eq:1}) in the asymptotic region
is approximated as
\begin{equation}
\bm{\Phi}''-\frac{1}{r^2}\pmatrix{2 & -1 & 0  \cr
                                  0 & 0  & 0 \cr
                                  0 & -1 & 2}{\bm \Phi}
+ \ldots =0 ,
\label{eq:su4i}
\end{equation}
where $\bm{\Phi}$ is a column vector
$\bm{\Phi}=\left(\Phi_1,\Phi_2,\Phi_3\right)^{T}$, and the dotted part
is multiplied by the powers of the exponentially decaying factor
$e^{-2sr}$.
Since the eigenvalues of the $3\times 3$ matrix in (\ref{eq:su4i}) are
$0$ and $2$ (doubly degenerate), there are two independent divergent
modes which behave as $r^2$ for large $r$.
In order to obtain a physically sensible $\Phi_m$, we have to adjust
two of the parameters of the solution to eliminate the divergent modes.
As we mentioned in sec. \ref{sec:neartheorigin}, the number of free
parameters in the $SU(4)$ solution $X$ regular at $r=0$ is three.
Therefore, we are left with only one free parameter.
Recalling that the other scalar $Y$ is already a
solution to the same differential equation as for $X$ with regular
behavior (\ref{eq:asY}) at $r\to\infty$, we find that $X$ must be
proportional to $Y$, implying that the solution is an uninteresting
one with parallel charge vectors.
Analysis for the case (ii) leads to the same
situation, namely, two divergent modes exist at infinity. Hence we
conclude that for (i) and (ii) there is no non-trivial regular
solution.

Next, let us consider the case (iii).
In this case, the magnetic charges of the solutions are $(-2,-2,1,3)$,
$(-3,0,0,3)$ and $(-3,-1,2,2)$ corresponding to three different
degeneracy patterns of $(y_a)$, $y_1=y_2$, $y_2=y_3$ and $y_3=y_4$,
respectively.
We shall take the case $y_1=y_2$.
Then, similarly to eq.\ (\ref{eq:su4i}), the differential equation
(\ref{eq:1}) is approximated for large $r$ as
\begin{equation}
\bm{\Phi}''-\frac{1}{r^2}\pmatrix{2 & -1 & 0 \cr
                                  0 &  0 & 0 \cr
                                  0 &  0 & 0}{\bm \Phi}
+ \ldots =0 .
\label{eq:su4iii}
\end{equation}
In this case, the eigenvalues of the $3\times 3$ matrix are $0$
(doubly degenerate) and $2$, and there is only one harmful
divergent mode.  Therefore, contrary to the cases (i) and (ii), there
can be non-trivial solutions since we still have two free
parameters in the solution after eliminating one divergent mode.

For these solutions regular at $r\to\infty$ we have
\begin{equation}
(X_1,Y_1) - (X_2,Y_2)=O\left(\frac{1}{r^2}\right) .
\label{eq:X1-X2}
\end{equation}
This is because eq.\ (\ref{eq:su4iii}) implies
\begin{equation}
\left(2\Phi_1-\Phi_2\right)'' -
\frac{2}{r^2}\left(2\Phi_1-\Phi_2\right)+ \ldots =0
\label{eq:diffeqX1-X2}
\end{equation}
for $2\Phi_1-\Phi_2=2r\left(X_1-X_2\right)$, and the divergent $r^2$
term in the solution at $r\to\infty$ has been eliminated by
fine-tuning to leave the $1/r$ term.
{}From eq.\ (\ref{eq:X1-X2}) and the asymptotic expressions
(\ref{eq:asX}) and (\ref{eq:asY}), we find that $(x_1,y_1)=(x_2,y_2)$
and $(e_1,g_1)=(e_2,g_2)$ hold for the present solution.
In the IIB string picture, two strings $1$ and $2$ degenerate
completely, and the situation is reduced to the three-pronged string.

\begin{figure}[htdp]
\begin{center}
\leavevmode
\epsfxsize=150mm
\epsfbox{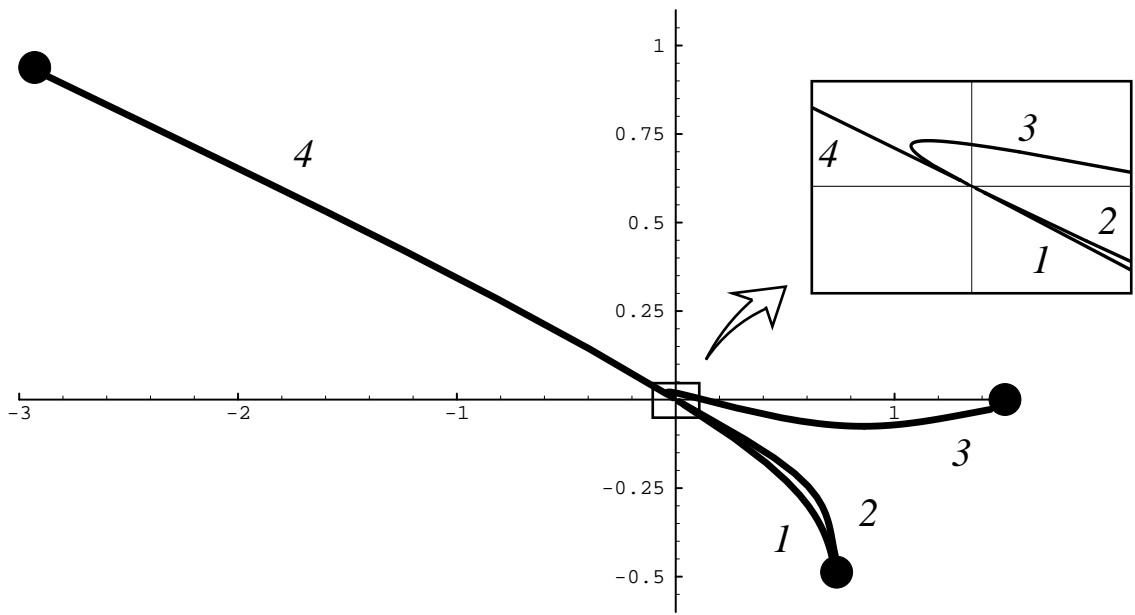}
\caption{A configuration with $(y_a)=(-1/2,-1/2,0,1)$. The locations
  of two D3-brane surfaces $1$ and $2$ at $r=\infty$
  coincide. The behavior near $r=0$ is magnified in the small window.
}
\label{fig:dege}
\end{center}
\end{figure}

An example of the solutions for $(y_a)=(-1/2,-1/2,0,1)$ obtained by a
numerical method is presented in fig.\ \ref{fig:dege}. We see that two
D3-brane surfaces $1$ and $2$ approaches to each other as $r\to\infty$
as mentioned above, though at the origin they behave in the way
explained in secs.\ \ref{sec:su(4)exact} and \ref{sec:neartheorigin}.

\section{The existence condition of a three-pronged string
and the force between two strings }
\label{sec:force}

\begin{figure}[htdp]
\begin{center}
\leavevmode
\epsfxsize=80mm
\epsfbox{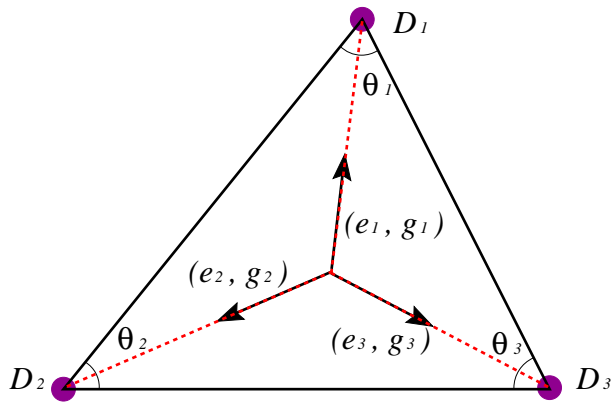}
\caption{
The triangle formed by three parallel D3-branes (solid lines).
The charge vectors are described by the lines with an arrow.
The dotted lines denote a three-pronged string connecting the
D3-branes.}
\label{fig:existcon}
\end{center}
\end{figure}

In the string picture,
the existence of a three-pronged string connecting three
parallel D3-branes depends on its two-form charges and
the relative locations of the D3-branes \cite{BER}.
Let us consider three D3-branes denoted by $D_1$, $D_2$ and $D_3$, and
a three-pronged string connecting them (fig.\ \ref{fig:existcon}).
We denote the two-form charges of the strings ending on the $D_1$,
$D_2$ and $D_3$ by $(e_1,g_1)$, $(e_2,g_2)$ and $(e_3,g_3)$,
respectively, where the charges are conserved at the junction:
$\sum_{i=1}^3 e_i=\sum_{i=1}^3 g_i=0$.
Since the force balance condition determines the relative directions
of the three strings meeting at the junction, the shape of
the triangle formed by the three D3-branes is constrained.
One can show easily that the necessary and sufficient
conditions for the existence of a three-pronged string connecting the
D3-branes are given by the following conditions for each angle
$\theta_i$ at the vertices $D_i$ of the triangle:
\be
\theta_i\le{\rm Angle}\left[ (e_j,g_j),(e_k,g_k)\right]
\ \ {\rm for\ different\ }i,j,k.
\label{angine}
\ee
Here Angle$[\bm{v}_1,\bm{v}_2]$ denotes the angle between the two
vectors $\bm{v}_1$ and $\bm{v}_2$ having a common initial point.
When the length of one of the strings vanishes, one of the equalities
holds in (\ref{angine}).
In such cases the configuration is a stable BPS state with two strings
located at the same point in the D3-brane world volume.
When the D3-brane locations are changed further out of
(\ref{angine}), the stability may be lost. Thus the process of
changing D3-brane locations
from (\ref{angine}) to the outside may be observed as
a decay process of a field theory BPS state of a three-pronged string
to two dyon states \cite{BER,FAY}.
On the other hand, when two strings with some two-form charges are
located at a finite distance in the D3-brane world volume, and
the configuration of the D3-branes satisfies (\ref{angine}),
the strings may attract each other and decay to the stable BPS state
of a three-pronged string.
Thus the evaluation of the force between two dyons would be another
way to see the conditions (\ref{angine}).
In this section, we will estimate the force between well-separated
two dyons following essentially the approach of \cite{FH,MAN},
and will actually reproduce (\ref{angine}).

\begin{figure}[htdp]
\begin{center}
\leavevmode
\epsfxsize=100mm
\epsfbox{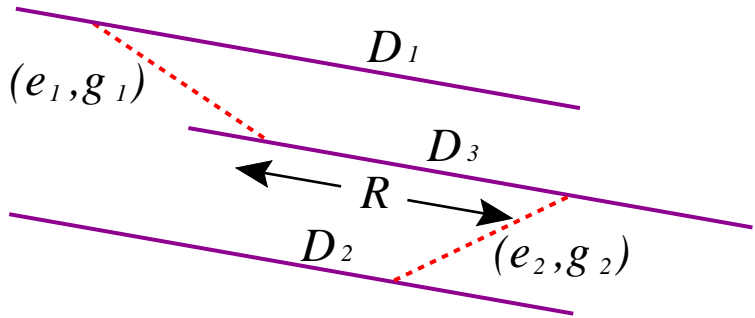}
\caption{
Two strings  with charges ($e_1,g_1$) and ($e_2,g_2$) (dotted lines)
connect one common and two different parallel D3-branes
(solid lines). The two strings are separated by $R$ in the D3-brane
world volume.}
\label{fig:exist2}
\end{center}
\end{figure}
Let us consider the following configurations of D3-branes and strings.
Three parallel D3-branes $D_1$, $D_2$ and $D_3$ are located at
$(x_1,y_1)$, $(x_2,y_2)$ and $(x_3,y_3)$ in the transverse
two-dimensional plane, respectively (fig.\ \ref{fig:exist2}).
An $(e_1,g_1)$ string connects $D_1$ and $D_3$, and an $(e_2,g_2)$
string connects $D_2$ and $D_3$. The two strings are separated by
a distance $R$ in the D3-brane world volume.
The two strings appear as two dyons with the electric and magnetic
charges $(e_1,g_1)$ and $(e_2,g_2)$ under the different $U(1)$
subgroups of $SU(3)$, respectively, in the D3-brane world volume
theory. The dyon corresponding to the $(e_1,g_1)$ string is denoted by
$d_1$ and the other one by $d_2$.
We assume the distance $R$ is large enough to allow treating
the dyons as point-like particles.

There act a Coulomb force between the two dyons.
The Coulomb potential is obtained as
\be
V_{\rm Coulomb}(R)=\frac{\pi}R (e_1e_2+g_1g_2) ,
\label{potcou}
\ee
by simply adding the electric and magnetic contributions.

Another force is generated by the long-range Higgs field.
Let us first evaluate the mass of the dyon $d_1$ by neglecting the
effect of the other dyon $d_2$.
By calculating $Q_{X,Y}$ and $M_{X,Y}$ from the asymptotic behaviors
of the electric and magnetic fields (\ref{eq:asX})-(\ref{eq:asB}),
and plugging them into (\ref{eq:BPSbound}),
the mass of the dyon $d_1$ is obtained as
\be
m_1=2\pi \sqrt{e_1^2+g_1^2}\ l_{13},
\label{mass2}
\ee
where $l_{13} =\sqrt{(x_1-x_3)^2+(y_1-y_3)^2}$
is the distance between $D_1$ and $D_3$.
This mass agrees with the corresponding string mass
by identifying $2\pi \sqrt{e_1^2+g_1^2}$ as the string tension.
We shall evaluate the potential energy between the dyons by
taking into account the $R$-dependence of the distance $l_{13}$
caused by the long-range effect of the other dyon $d_2$.

{}From (\ref{eq:asX}) and (\ref{eq:asY}), the long-range behaviors of
the Higgs field generated by the dyon $d_2$ are given by
\bea
&&X\sim\diag(x_1,x_2,x_3) + \frac1{2R}\diag(0,u_2,u_3),
\nn\\
&&Y\sim\diag(y_1,y_2,y_3) + \frac1{2R}\diag(0,v_2,v_3),
\label{lonhig}
\eea
where the $u_a$ and $v_a$ are related to the charges by
(\ref{eq:(u,v)}).
The $R$-dependent terms of the third components cause the
$R$-dependence of the distance $l_{13}$.
The factors are expressed explicitly as
\be
(u_3,v_3)=\sqrt{e_2^2+g_2^2} \ \widehat{\bm{l}}_{23},
\label{u3v3}
\ee
where $\widehat{\bm{l}}_{23}$ is a unit vector parallel to the string
connecting $D_2$ and $D_3$:
\be
\widehat{\bm{l}}_{23}
=\frac{(x_2-x_3,y_2-y_3)}{\sqrt{(x_2-x_3)^2+(y_2-y_3)^2}}.
\label{l23}
\ee
Plugging these asymptotic behaviors into (\ref{mass2}) and
taking the first order in $1/R$, we obtain
the potential energy from the long-range Higgs field as
\be
V_{\rm Higgs}(R)=-\frac\pi{R}\sqrt{e_1^2+g_1^2}\sqrt{e_2^2+g_2^2}
\ \widehat{\bm{l}}_{13}\cdot\widehat{\bm{l}}_{23} .
\label{pothig}
\ee
Adding this to the Coulomb potential (\ref{potcou}), the total
potential energy is obtained as
\be
V(R)=-\frac\pi{R}\left( \sqrt{e_1^2+g_1^2}\sqrt{e_2^2+g_2^2}
\ \widehat{\bm{l}}_{13}\cdot\widehat{\bm{l}}_{23}
- e_1 e_2 - g_1 g_2\right).
\ee

The cancellation between the Coulomb and the Higgs contributions
occur when the angle between $\widehat{\bm{l}}_{13}$ and
$\widehat{\bm{l}}_{23}$ equals that between the two vectors
$(e_1,g_1)$ and $(e_2,g_2)$.
In case the former angle is smaller than the latter, the force between
the two dyons is attractive, and vice versa.
Thus we obtain the inequality for $\theta_3$ in (\ref{angine})
from a field theoretical viewpoint.
Repeating similar discussions for the other combinations of two
strings, we obtain all the inequalities in (\ref{angine}).


\section{Summary and discussions}
\label{sec:summary}

In this paper, we have constructed 1/4 BPS states
in ${\cal N}=4$ $SU(N)$ SYM theory by solving explicitly the
Bogomol'nyi equations under the assumption of spherical symmetry.
The solutions correspond to the string theory BPS states of a
multi-pronged string connecting $N$ different parallel D3-branes.
Here each string ending on D3-branes carries the two-form charges
$(e_1,N-1),(e_2,N-3),\cdots,(e_N,-N+1)$, where $e_a$ take real
values satisfying $\sum_a e_a=0$.
The NS-NS charges $e_a$ (or electric charges in the SYM theory) may be
quantized if the quantization of the collective modes
of our solutions is performed.
We have also shown that, by fine-tuning some of the parameters,
we can construct solutions with magnetic charges different from above.

As we have studied in secs.\ \ref{sec:string-network}, \ref{sec:shape}
and appendix \ref{app:network}, the trajectories of D3-branes have
interesting behaviors.
We have shown that, from the asymptotic behaviors of our solutions,
we can generate the BPS saturated configuration of the corresponding
multi-pronged string, that is,  straight
strings compose a tree diagram with junctions and the forces balance
at each junction.
On the other hand, the behaviors at finite $r$ are
quite different from the above IIB string picture.
The trajectories bend non-trivially and all the D3-brane surfaces
meet at the origin $r=0$.
Nonetheless, an interesting feature of the trajectories is that we can
actually generate the string picture configuration at every $r$ by
defining appropriately the effective electric and magnetic charges.
At the origin $r=0$ the directions of all the charge vectors with
non-vanishing magnetic charges degenerate,
and hence all the D3-brane trajectories with non-vanishing R-R
two-form string charges meet there with a common tangent.
This feature apparently violates the $SL(2,Z)$ duality symmetry, and
may come from our technical preference that we treat the SYM theory
merely classically, i.e. electrically \cite{ours}.

We have constructed exact solutions for general $SU(N)$
in the case the vacuum expectation values of one of the Higgs fields
($Y$) are parallel to the magnetic charges.
The number of the free parameters of our exact solutions
are given by $N$, one from the overall
factor of the Higgs field and $N-1$ from the vacuum expectation
values of another Higgs field ($X$).
In the general cases of our (non-exact) solutions,
the vacuum expectation values of the first Higgs field can be varied
continuously,
and hence the total number of the free parameters is expected to be
given by $2N-2$.
The solutions should be regular at the whole space including $r=0$ and
should converge to finite values at $r\rightarrow\infty$.
We have checked the number of the free parameters up to $SU(6)$
by expanding explicitly the second-order differential equations
(\ref{eq:1}) at $r\sim 0$ and checking numerically the safe
convergence of the solutions at $r\to\infty$. A general proof of this
fact has not been given.

This number of the free parameters of our solutions is interesting
because it  does not agree with the expectation in the IIB string
picture, as discussed in our previous work \cite{ours}.
Let us consider a tree diagram composed of straight string segments
and three-string junctions in a two-dimensional plane.  Here we assume
the number of external strings ending on D3-branes is $N$ and
they have the same fixed R-R charges as the magnetic charges of our
solutions.
We have the freedom to take $N-1$ NS-NS charges (the other one is
determined by the charge conservation). Then the relative directions
of the strings in the diagram are determined by the force balance
condition. We have the freedom to take the $2N-3$ lengths of each
string segment in the diagram. We should not count the freedom of the
parallel shift and the rotation of the whole diagram, because we fixed
these  degrees of freedom in obtaining our solutions (see section
\ref{sec:sph_symm}).
Thus the total number of the degrees of freedom is given by
$3N-4$, which exceeds the above degrees of freedom of our solutions by
$N-2$.
We do not have any reliable explanation of this discrepancy, but may
point out that the freedom of choosing the two-form (or electric and
magnetic) charges does not seem to
be fully incorporated as can be seen from the $r\sim0$ behaviors of
our solutions discussed in the second paragraph of this section.
Thus we hope the number of free parameters of such solutions  may
agree with the string picture if we find a better formulation in field
theory respecting the $SL(2,Z)$ duality symmetry to describe the 1/4
BPS states.

As pointed out in section \ref{sec:su(4)exact},
we found two different multi-pronged string configurations which
can correspond to the same field theory BPS states
in certain parameter regions of our solutions.
This may question the one-to-one correspondence between the BPS states
in string theory and those in field theory, and the
resolution of this fact may be open.

In section \ref{sec:force},
we discussed the force between two strings connecting
one common and two different parallel D3-branes by regarding the strings
as dyons and
calculating the long-range force between them in the SYM theory.
In a certain parameter region of the D3-brane coordinates, the force is
attractive, and a three-pronged string can be formed as a
bound state of the two strings. This result is another support to the
existence of a three-pronged string connecting D3-branes.

The supermultiplet of a 1/4 BPS state should contain states with
spins higher than one (see appendix \ref{sec:susy}).
The appearance of a field with such a high spin is unusual
in a field theory without gravity. It is intriguing to note that
a 1/4 BPS state is a non-local object and cannot appear as
an elementary local field.
To see this, let us consider a $(0,1)$ string
connecting two parallel D3-branes separated by a distance $l$.
This string state corresponds to a monopole in $SU(2)$ SYM theory.
One sees that the energy of a monopole is given by
$E_{(0,1)}=Tl/g$, while its width is
$\delta r_{(0,1)}=(Tl)^{-1}$ by viewing the explicit expression
of the monopole configuration, where $T$ and $g$ are the string
tension and the string coupling constant, respectively \cite{Hashi}.
By performing the $SL(2,Z)$ duality transformation on
the equation $E_{(0,1)} \delta r_{(0,1)}=1/g$,
we obtain $E_{(p,q)}\delta r_{(p,q)}=|p+q\tau|^2/{\rm Im}\tau$
for a $(p,q)$ string in general.
Thus one cannot take $\delta r\rightarrow 0$ limit simultaneously
for two strings with non-parallel NS-NS and R-R charges
by taking an appropriate limit of $\tau$.
This fact implies that a 1/4 BPS state cannot appear as an
elementary local field in any $SL(2,Z)$ equivalent description,
because it has non-parallel electric and magnetic charges.

In appendix \ref{app:BornInfeld},
we have shown that our solutions are also solutions of the non-abelian
BI action proposed in \cite{Tseytlin}.
Among the developments of the BI physics
\cite{CH,Hashi,Gibb,LPT,BI}, a remarkable fact is the coincidence
between the brane worldvolume
approach and the target space approach \cite{LPT}. Since any solution
of supergravity corresponding to string junctions has not
been constructed yet, we hope our brane realization of string
junctions will give a certain insight to a supergravity
realization of string junctions.

It is clear that there are some unsatisfactory points to be overcome
in our formulation to describe the general 1/4 BPS states.
We need more freedom to choose the magnetic charges.
A BPS state of multi-pronged string can have loops \cite{SENNET,KOL},
but how to incorporate loops in our formulation is not clear.
An $SL(2,Z)$ invariant formulation is also needed.
It is known that Nahm equations \cite{NAH} describe monopoles on
D3-branes by 1-brane dynamics \cite{DIA}
rather than the D3-brane dynamics as we did.
It might shed light on the above problems
if the formulation is extended to
multi-pronged strings considering appropriate boundary conditions of
1-brane fields at junctions like those given in \cite{CALTHO}.

Since we use only two adjoint Higgs fields in our strategy of solving
the Bogomol'nyi equations,
our results may be also applicable to ${\cal N}=2$ SQCD.
Thus the states discussed in this paper might be related to the exotic
states with non-parallel electric and magnetic charges
observed recently in \cite{sugimoto}
in ${\cal N}=2$ MQCD \cite{BRANE}
(See also \cite{FH,AHAYAN} for some discussions on such exotic states).

Note added: While this paper was in the final stage of preparation,
an article \cite{kawano} appeared which has an overlap with our
discussions on exact solutions.

\vspace{.5cm}
\noindent
{\large\bf Acknowledgments}\\[.2cm]
We would like to thank I.\ Kishimoto, Y.\ Murakami, S.\ Sugimoto
for valuable discussions and comments, and S.\ Sugimoto also for
explaining his results in \cite{sugimoto}.
N.S. would like to thank H.\ Kunitomo and T.\ Nakatsu for stimulating
comments at the Sapporo Winter School in Niseko in 1998.

\clearpage
\appendix
\vspace{1.5cm}
\centerline{\Large\bf Appendix}
\appendix

\section{String networks in the IIB picture}
\label{app:network}

In this appendix, we present a proof of the
properties (A) and (B) of sec.\ \ref{sec:string-network} concerning
the presence of the string networks for general $N$.
The following proof is based on a substitution rule which allows us to
reduce the number $N$ of the strings emerging from the D3-branes (we
call these strings ``external string'' hereafter) by one.

For any two among the $N$ external strings,
say the strings $1$ and $2$, let us consider a new string $12$
carrying the charge $(e_{12},g_{12})=(e_1+e_2,g_1+g_2)$.
The transverse coordinate $(x_{12},y_{12})$ from which the new string
$12$ emerges is determined by the conditions that the BPS bound
$\EBPS$ (\ref{eq:BPSbound}) of the energy and the angle $\theta$
of eq.\ (\ref{eq:theta}) are common in both the original system of $N$
external strings $(1,2,3,\ldots,N)$ and the new system of $N-1$ external
strings $(12,3,\ldots,N)$. Then, this new string $12$ satisfies the
two properties:
\begin{enumerate}
\item[(I)]
The junction point of the strings $1$ and $2$ lies on the
string $12$.
\item[(II)]
Taking this junction point as the common end of the three strings $1$, 
$2$ and $12$, the mass of the string $12$ is equal to the sum of the
masses of the strings $1$ and $2$.
\end{enumerate}

To show the properties (I) and (II), note first that the condition
determining $(x_{12},y_{12})$ is expressed as
\begin{equation}
C_{12}\vxy{12} = C_1\vxy{1} + C_2\vxy{2} ,
\label{eq:xy12}
\end{equation}
where $C_a$ ($a=1,2,12$) is the $2\times 2$ matrix given by
\begin{equation}
C_a=\pmatrix{g_a & -e_a\cr e_a & g_a} .
\label{eq:C}
\end{equation}
Then, the property (I) claims that, for $t_1$ and $t_2$ determined by
the first equality of
\begin{equation}
\vxy{1}+t_1\vuv{1}=\vxy{2}+t_2\vuv{2}=\vxy{12}+t_{12}\vuv{12} ,
\label{eq:t12}
\end{equation}
there exits $t_{12}$ which satisfies the second equality.
Such $t_{12}$ is easily found by eliminating $(x_a,y_a)$ ($a=1,2$)
from eq.\ (\ref{eq:xy12}) using eq.\ (\ref{eq:t12}) to get
\begin{equation}
t_{12}C_{12}\vuv{12} = t_1 C_1\vuv{1} + t_2 C_2\vuv{2} ,
\label{eq:solvet12}
\end{equation}
which is equivalent to
\begin{equation}
t_{12}\left(e_{12}^2 + g_{12}^2\right)
= t_{1}\left(e_{1}^2 + g_{1}^2\right)
+ t_{2}\left(e_{2}^2 + g_{2}^2\right) .
\label{eq:t12t1t2}
\end{equation}
This equation (\ref{eq:t12t1t2}) also implies the property (II)
since the tension and the length of the strings are given by
$T_a=2\pi\sgn(t_a)\sqrt{e_a^2+g_a^2}$ and
$\ell_a=|t_a|\sqrt{e_a^2+g_a^2}$, respectively:
\begin{equation}
T_{12}\ell_{12}=T_{1}\ell_{1} + T_{2}\ell_{2} .
\label{eq:sum_mass12}
\end{equation}

Now, the properties (I) and (II) implies that we can reduce the
problem of proving (A) and (B) for $N$ external strings to that for
$N-1$ external strings. Since we know that (A) and (B) hold for $N=3$,
we conclude that they are valid for an arbitrary $N$.


\section{Solving the eigenvalue problem (\protect\ref{eq:comp})}
\label{app:v}

In this appendix, we shall solve the eigenvalue problem
(\ref{eq:comp}), namely, we obtain the eigenvalues $p(p+1)$ and the
corresponding eigenvector $v_m^{(p)}$.
For this purpose let us introduce a polynomial $V^{(p)}(z)$ of a
variable $z$ as
\begin{eqnarray}
\label{eq:difi}
V^{(p)}(z)\equiv \sum_{m=1}^{N-1} v_m^{(p)} z^m.
\end{eqnarray}
Then, eq.\ (\ref{eq:comp}) is expressed as a second order
differential equation for $V^{(p)}(z)$:
\begin{eqnarray}\label{eq:diff}
&&z(z-1)^2 V^{(p)\,\prime\prime}
+(z-1)\Bigl[N+1 - (N-3)z \Bigr]V^{(p)\,\prime}
\nn\\
&&\hspace{35mm}+
\left[\frac{N+1}{z} - (N-1)z - p(p+1)\right]V^{(p)}=0 ,
\label{eq:diffeqV}
\end{eqnarray}
where the prime denotes the differentiation with respect to $z$.
In order to transform the differential equation (\ref{eq:diffeqV})
into a more familiar one, we rewrite it in terms of $\wt{V}^{(p)}(z)$
defined by
\begin{equation}
V^{(p)}(z)= z(1-z)^{p-1}\wt{V}^{(p)}(z) .
\end{equation}
The differential equation for $\wt{V}^{(p)}$ reads
\begin{eqnarray}
&&z(1-z)\wt{V}^{(p)\,\prime\prime} +
\Bigl[-N+1-(2p-N+3)z\Bigr]\wt{V}^{(p)\,\prime}
\nn\\
&&\hspace{55mm}
-(p-N+1)(p+1)\wt{V}^{(p)}=0 .
\label{eq:diffeqW}
\end{eqnarray}
This hypergeometric differential equation has a polynomial solution
when $p$ takes integer values $p=1,2,\ldots,N-1$, and we have
\begin{eqnarray}
V^{(p)}(z)=z(1-z)^{p-1}\sum_{k=0}^{N-p-1}
\frac{\C{N-p-1}{k} \;\C{k+p}{k}}{\C{N-1}{k}}\,z^k .
\label{eq:Vfinal}
\end{eqnarray}
The eigenvector $v_m^{(p)}$ can be read off from (\ref{eq:difi}) and
(\ref{eq:Vfinal}).
For example, $v_m^{(1)}$ and $v_m^{(2)}$ are given (up to the
normalization constant) by
\begin{eqnarray}
&&v_m^{(1)} = m\overline{m} , \label{eq:vm1}\\
&&v_m^{(2)} = m\overline{m}\left(m -\overline{m}\right) ,
\end{eqnarray}
with $\overline{m}\equiv N-m$.

\section{The coefficient $c_p$}
\label{app:c_p}

In this appendix we derive the solution of (\ref{eq:Gauss-f}) in an
infinite Taylor series at $y=1$ and obtain the coefficient $c_p$ in
(\ref{eq:c_p}) by
an analytic continuation to the expression (\ref{eq:solution}).

We should find the solutions $\wt{\varphi}^{(p)}$ of
(\ref{eq:Gauss-f}) which damps faster than $(1-y)^p$ at $y\sim 1$.
Although the solution of the hypergeometric differential equation
(\ref{eq:Gauss-f})
which is regular at $y=1$ may be expressed formally by the
hypergeometric series $F(-p,-p,-2p;1-y)$, this does not make sense
because of a zero in the denominators.
Thus the real solution may be obtained by taking an
appropriate limit of $\ep_i\rightarrow0$ in the expression
$F(-p+\ep_1,-p+\ep_2,-2p+\ep_3;1-y)$.
The higher terms from order $(1-y)^{2p+1}$ of this series
have a $1/\ep_3$ singularity at $\ep_3=0$, and
the series of these singular terms\footnote{Hence this series damps
faster than $(1-y)^p$ as desired.}
can be shown to satisfy by itself a hypergeometric differential
equation which converges to (\ref{eq:Gauss-f}) in the limit
$\ep_i\rightarrow0$.
Since this partial series
$\lim_{\ep_3\rightarrow0} \ep_3 F(-p+\ep_1,-p+\ep_2,-2p+\ep_3;1-y)$
behaves like $\ep_1\ep_2$ at $\ep_{1,2}\rightarrow0$,
we cancel this factor and obtain a solution
\be
\wt{\varphi}^{(p)}=
\lim_{\ep_{1,2}\rightarrow0} \lim_{\ep_{3}\rightarrow0}
\frac{\ep_3}{\ep_1\ep_2} F(-p+\ep_1,-p+\ep_2,-2p+\ep_3;1-y),
\label{solvarphi}
\ee
where we wrote explicitly that the limit $\ep_{3}\rightarrow0$ should
be taken first.

A well-known formula for an analytic continuation of the
hypergeometric function reads
\bea
&&F(\alpha,\beta,\gamma;1-y)=
\frac{\Gamma(\gamma)\Gamma(\gamma-\alpha-\beta)}
{\Gamma(\gamma-\alpha)\Gamma(\gamma-\beta)}
F(\alpha,\beta,\alpha+\beta-\gamma+1;y)
\nn\\
&&\qquad\qquad\qquad
+y^{\gamma-\alpha-\beta}
\frac{\Gamma(\gamma)\Gamma(\alpha+\beta-\gamma)}
{\Gamma(\alpha)\Gamma(\beta)}
F(\gamma-\alpha,\gamma-\beta,\gamma-\alpha-\beta+1;y).
\label{anacon}
\eea
We also use a well-known expansion formula for the gamma function:
\be
\Gamma(\ep-n)=\frac{(-1)^n}{n!}\left(\frac1\ep+\sum_{l=1}^n
\frac1l+\psi(1)+O(\ep)\right),
\label{expgam}
\ee
where $\psi(1)$ is the Euler constant. Taking the limit in
(\ref{solvarphi}) by using (\ref{anacon}) and (\ref{expgam}), we obtain
\be
\wt{\varphi}^{(p)}=-\frac{(p!)^2}{(2p)!}
\left(2\left(\sum_{l=1}^p\frac1l\right)
F(-p,-p,1;y) + F^*(-p,-p,1;y)+\ln(y)F(-p,-p,1;y)
\right),
\ee
where we have used the definition
$F^*(\alpha,\beta,\gamma;z)\equiv
\left(\frac\partial{\partial\alpha}+\frac\partial{\partial\beta}
+2\frac\partial{\partial\gamma}\right)F(\alpha,\beta,\gamma;z)$.
This is the solution (\ref{eq:solution})
with $c_p=1/(2\sum_{l=1}^p\frac1l)$.


\section{Supersymmetric aspects of the solution}
\label{sec:susy}

It was argued in \cite{FH,BER} that the solution with non-parallel
electric and magnetic charges has $1/4$ supersymmetry.
In this appendix we shall examine this property in detail.

The Lagrangian of the 3+1 dimensional
${\cal N} =4$ supersymmetric Yang-Mills theory
is given by
\cite{N=4}
\begin{eqnarray}
&&{\cal L} = -\frac{1}{4}F_{\mu\nu}^aF^{\mu\nu a} +
\frac{i}{2}\overline{\lambda}^a_K\gamma^\mu  D_\mu\lambda^a_K
+\Half (D_\mu A_i^a)^2
+\Half (D_\mu B_i^a)^2
\nn\\
&&\qquad\qquad
-\frac{i}{2}gf_{abc}\overline{\lambda}^a_K
(\alpha^i_{KL}A_i^b + i\beta^i_{KL}\gamma_5 B_i^b)\lambda^c_L
\nn\\
&&\qquad\qquad
-\frac{1}{4}g^2\left(
  \left( f_{abc}A_i^b A_j^c \right)^2
+  \left( f_{abc}B_i^b B_j^c \right)^2
+ 2 \left( f_{abc}A_i^b B_j^c \right)^2
\right),
\end{eqnarray}
where $K$ and $L$ are the indices of the {\bm 4} representation of
the $SU(4)$ $R$-symmetry.
The matrices $\alpha$ and $\beta$ span the antisymmetric part of
${\bm 4}\times{\bm 4}$, which transform in {\bm 6}:
\begin{eqnarray}
&&\alpha_1=\pmatrix{ & \sigma_1\cr -\sigma_1 & } ,
\qquad
\alpha_2=\pmatrix{  & \sigma_3\cr -\sigma_3 & } ,
\qquad
\alpha_3=\pmatrix{i \sigma_2& \cr & i\sigma_2 } ,
\nn\\[5pt]
&&\beta_1=\pmatrix{ &i \sigma_2\cr i\sigma_2 & } ,
\qquad
\beta_2=\pmatrix{ & {\bm 1}\cr -{\bm 1} & } ,
\qquad
\beta_3=\pmatrix{-i \sigma_2& \cr & i\sigma_2 } .
\end{eqnarray}
Here $\sigma_i$ denote the Pauli matrices. The Lagrangian is invariant
under the ${\cal N}=4$ supersymmetry transformation. The supersymmetry
transformation of the gaugino is given by
\begin{eqnarray}
&&\delta\lambda_K=
\left[
  \frac{-i}{2}\sigma^{\mu\nu}F_{\mu\nu} -i\gamma^\mu D_\mu
    \left(\alpha^i A_i + i\beta^i \gamma_5 B_i \right)
\right.
\nn\\
&&\qquad\qquad
\left.
+\frac{i}{2}g\epsilon_{ijk}\alpha^k [A^i,A^j]
+\frac{i}{2}g\epsilon_{ijk}\beta^k [B^i,B^j]
+g\alpha^i\beta^j [A^i,B^j]\gamma_5
\right]_{KL}
\epsilon_L,
\label{susytran}
\end{eqnarray}
where the parameter $\epsilon_L$ denotes a Majorana spinor.

In investigating the supersymmetric aspects of our solutions, we
identify $(A_3,B_3)$ with $(X,Y)$ and put $A_{1,2}$ and $B_{1,2}$
equal to zero due to the $SO(6)$ ($\sim SU(4)$) $R$-symmetry of the
action. Plugging the Bogomol'nyi equations (\ref{eq:D0X=0}),
(\ref{eq:Dflat}), (\ref{eq:DX=E}) and (\ref{eq:DY=B}) into the above
supersymmetry transformation (\ref{susytran}), we obtain
\begin{eqnarray}
\delta\lambda^a_K=
\left[
  -i(\sigma^{0i}\delta_{KL}+\gamma^i\alpha^3_{KL}){\cal E}^a_i
  -\left(\frac{i}{2}\sigma^{ij}\epsilon_{ijk}\delta_{KL}
  - \gamma_k\gamma_5\beta^3_{KL}\right) {\cal B}^{ka}
\right]
\epsilon_L.
\end{eqnarray}
In our solutions the electric and magnetic fields behaves
differently from each other, and thus we obtain two
constraints\footnote{Our convention of the Dirac matrices are
$$
\gamma^0 =\pmatrix{1 & \cr & -1} ,
\quad
\gamma^i =\pmatrix{ &\sigma_i \cr -\sigma_i & },
\quad
\gamma_5=\pmatrix{ &1 \cr 1 & } ,
\quad
\sigma^{\mu\nu}=\frac{i}{2}[\gamma^\mu,\gamma^\nu].
$$
} on $\epsilon_L$:
\begin{eqnarray}
\label{eq:e1}
&&\left[
\delta_{KL}\pmatrix{0& \sigma_i \cr \sigma_i &0}
-i \pmatrix{i\sigma_2&0 \cr 0&i\sigma_2 }_{KL}
\pmatrix{0& \sigma_i \cr -\sigma_i &0}
\right]\epsilon_L=0,
\nn\\[5pt]
&&\left[
\delta_{KL}\pmatrix{\sigma_i&0 \cr 0&\sigma_i }
+i \pmatrix{-i\sigma_2&0 \cr 0&i\sigma_2 }_{KL}
\pmatrix{ \sigma_i&0 \cr 0&-\sigma_i }
\right]\epsilon_L=0.
\end{eqnarray}
Representing the Majorana spinor $\epsilon_L$ by a Weyl spinor $\eta_L$
\begin{eqnarray}
\label{eq:e_L}
\epsilon_L = \pmatrix{i\sigma_2\eta_L^*\cr \eta_L} ,
\end{eqnarray}
and substituting (\ref{eq:e_L}) into (\ref{eq:e1}), we find the
conditions
\begin{eqnarray}
\pmatrix{1&i& & \cr -i & 1 & & \cr & & 1& -i\cr & & i & 1}
\pmatrix{\eta_1\cr \eta_2\cr \eta_3\cr \eta_4}
=\pmatrix{1&-i& & \cr i & 1 & & \cr & & 1& -i\cr & & i & 1}
\pmatrix{\eta_1\cr \eta_2\cr \eta_3\cr \eta_4}
=0 .
\end{eqnarray}
The solution is obtained as
\begin{eqnarray}
\label{eq:comb}
\eta_L=a_L \eta, \qquad
(a_1,a_2,a_3,a_4)= e^{i\theta_R}\left(0,0,i,1\right),
\end{eqnarray}
where $\eta$ is a constant Weyl spinor.
Thus the BPS states discussed in this paper preserve one
supersymmetry (\ref{eq:comb}) out of the original four, and the phase
factor in eq.\ (\ref{eq:comb}) indicates the remaining $U(1)$
$R$-symmetry.

In this ${\cal N}=4$ system, we have two independent central charges
$Q_X$ and $M_Y$ after an appropriate $SO(6)$ $R$-transformation
(a subgroup of this transformation is used to put $\theta$ to zero
in sec\ \ref{sec:SU(N)}).
The formula for the lower bound of the energy is given by
\begin{eqnarray}
&&U=\int\! d^3x\; \Half\tr\left\{
\left(\E_i\!\mp\! D_i\X\right)^2
+\left(\B_i\!\mp\! D_i\Y\right)^2
+\left(D_0 \X\right)^2 +\left(D_0 \Y\right)^2 -\com{X}{Y}^2
\right\}
\pm Q_\X \pm M_\Y
\nn\\
&&\phantom{H}
\geq \mbox{max}\Bigl\{\;|Q_\X+M_\Y|\;,\;|Q_\X-M_\Y|\;\Bigr\},
\label{eq:appBPSbound}
\end{eqnarray}
which depends on the relative sign of the charges $Q_X$ and $M_Y$.
Though we solve only one of the two cases in this paper, this is not
essential because the Bogomol'nyi equations of another case are just
obtained by the substitution $X\rightarrow -X$.
Since our solutions have non-zero $Q_X$ and $M_Y$, their masses
saturate either $|\,Q_X+M_Y|$ or $|\,Q_X-M_Y|$, and not both.
Generally, states saturating $k$ bounds are generated by
$2({\cal N}-k)$ fermionic creation operators made of the supercharges.
Hence our BPS configuration of SYM belongs to a
supermultiplet with $n\times 2^6$ ($n$: integer) components.
This implies that this multiplet contains states with spins higher
than or equal to $3/2$.


\section{The non-Abelian Born-Infeld action}
\label{app:BornInfeld}

We shall show in this appendix that the BPS saturated solutions in
this and our previous papers \cite{ours} are also solutions of the
non-abelian Born-Infeld (BI) action proposed in \cite{Tseytlin}.
We will follow the discussions given in \cite{Hashi}, where
the usual monopole and dyon solutions in $SU(2)$ SYM theory were
shown to satisfy the equations of motion from the non-abelian BI
action.

The ordinary D3-brane effective action is given by the abelian BI
action
\begin{equation}
S = \int\! d^4x\; \sqrt{-\det (g_{\mu\nu}+F_{\mu\nu})},
\label{abeact}
\end{equation}
where we have ignored the overall numerical factor, which
is irrelevant in the discussions here,
and have appropriately normalized the gauge fields
to be consistent with eq.\ (\ref{eq:H}).
In our solutions, the D3-branes fluctuate in the transverse
two dimensional plane with coordinates $X$ and $Y$.
Hence the induced metric $g_{\mu\nu}$ on the D3-branes is given by
\begin{eqnarray}
g_{\mu\nu}=\eta_{\mu\nu}+\p_\mu X \p_\nu X + \p_\mu Y \p_\nu Y,
\label{indmet}
\end{eqnarray}
where $\eta_{\mu\nu}$ denotes the flat Minkowski metric.
The action (\ref{abeact}) with the metric (\ref{indmet}) can be simply
deduced \cite{Gibb} by the dimensional reduction
from the $6$-dimensional BI action with a flat metric:
\begin{eqnarray}
  S = \int\! d^6x\;\sqrt{-\det (\eta_{ab}+F_{ab})}.
\label{act6}
\end{eqnarray}
Here the fluctuations of the four-dimensional induced metric
(\ref{indmet}) are
incorporated in the six-dimensional gauge fields in (\ref{act6}) by
identifying $X$ and $Y$ with some  components through T-duality:
$X=A_4, Y=A_5$.

A non-abelian version of the BI action suffers from the ordering
ambiguities of the non-commutative field components. This ambiguity is
fixed in the proposal of \cite{Tseytlin} by taking the symmetrized
trace operation STr. The action is defined by
\begin{eqnarray}
\label{eq:nbi}
    S = \int\! d^6x\; {\rm STr} \sqrt{-\det (\eta_{ab}+F_{ab})},
\end{eqnarray}
where the $6$-dimensional field strengths are given by
\begin{eqnarray}
F_{ab}=\pmatrix{
0&{\cal E}_1&{\cal E}_2&{\cal E}_3&D_0X&D_0Y\cr
-{\cal E}_1&0&{\cal B}_3&-{\cal B}_2&D_1X&D_1Y\cr
-{\cal E}_2&-{\cal B}_3&0&{\cal B}_1&D_2X&D_2Y\cr
-{\cal E}_3&{\cal B}_2&-{\cal B}_1&0&D_3X&D_3Y\cr
-D_0X&-D_1X&-D_2X&-D_3X&0&-i[X,Y] \cr
-D_0Y&-D_1Y&-D_2Y&-D_3Y&i[X,Y]&0 }
\label{fiestr}
\end{eqnarray}

Since the part inside the symmetrized trace in eq.\ (\ref{eq:nbi}) can
be expanded in a polynomial form in $F_{ab}$, we can treat this part
as if $F_{ab}$ were an abelian variable before symmetrization.
The variation of the action gives
\begin{equation}
\delta S = \Half\,{\rm STr}
\left[\left\{\sqrt{-\det(\eta_{cd}+F_{cd})}\;
 (\eta+F)^{-1}_{ab}
\right\}\delta F^{ba}
\right].
\label{delta}
\end{equation}
In evaluating (\ref{delta}) for our configurations, we are allowed to
substitute eqs.\ (\ref{eq:D0X=0}), (\ref{eq:Dflat}), (\ref{eq:DX=E})
and (\ref{eq:DY=B}) into the curly bracket part $\{\cdots\}$ before
the symmetrizing operation.
This is because these four equations are linear relations among the
components of $F_{ab}$. Therefore, after a little algebra, the
quantity inside the curly bracket of (\ref{delta}) is simply replaced
by $-F_{ab}$.
Hence we find a similar situation as \cite{Hashi},
that is, there is no need to distinguish the symmetrized
trace operation from the ordinary one, since we have only two
non-commutative elements in the symmetrized trace of
(\ref{delta}). Noticing that $\delta F_{ab}=D_{[a} \delta A_{b]}$ and
performing an integration by parts, the equations of motion are
reduced to
\begin{eqnarray}
\label{eq:YMeq}
  D^aF_{ab}=0,
\end{eqnarray}
which are just the ones from the SYM action. The BPS conditions
(\ref{eq:D0X=0}), (\ref{eq:Dflat}), (\ref{eq:DX=E}) and (\ref{eq:DY=B})
satisfy eq.\ (\ref{eq:YMeq}). Therefore we have shown that our BPS
saturated configuration is also a solution of the equations of motion
from the non-abelian BI action.

\newcommand{\J}[4]{{\sl #1} {\bf #2} (#3) #4}
\newcommand{\andJ}[3]{{\bf #1} (#2) #3}
\newcommand{\AP}{Ann.\ Phys.\ (N.Y.)}
\newcommand{\MPL}{Mod.\ Phys.\ Lett.}
\newcommand{\NP}{Nucl.\ Phys.}
\newcommand{\PL}{Phys.\ Lett.}
\newcommand{\PR}{Phys.\ Rev.}
\newcommand{\PRL}{Phys.\ Rev.\ Lett.}
\newcommand{\PTP}{Prog.\ Theor.\ Phys.}
\newcommand{\hep}[1]{{\tt hep-th/{#1}}}

\end{document}